\documentclass[a4paper,11pt]{article}
\usepackage{jheppub} 
\usepackage{lineno}

\arxivnumber{2312.16660} 

\newif\ifTwocolumn
\Twocolumnfalse

\makeatletter
\AtBeginDocument{
  \if@twocolumn
    \Twocolumntrue
  \else
    \Twocolumnfalse
  \fi
}
\makeatother

\usepackage{amsfonts, amssymb, amsmath, amsthm, mathtools,wasysym}
\usepackage{breqn}
\usepackage{physics}
\usepackage{graphicx}
\usepackage{xcolor}
\usepackage[caption=false]{subfig}
\newcommand{\phantomsubcaption}[1]{
    {
        \captionsetup[subfloat]{farskip=0pt,captionskip=0pt}
        \captionsetup[subfigure]{labelformat=empty}
        \subfloat{#1}
    }%
}
\usepackage{hyperref}
\hypersetup{
    colorlinks=true,
    linkcolor=blue,
    filecolor=magenta,      
    urlcolor=cyan,
    }
\usepackage{cleveref}
\newcommand{\onlinecite}[1]{\hspace{-1 ex} \nocite{#1}\citenum{#1}}
\usepackage{indentfirst}
\usepackage{tikz}

\title{
Unveiling chiral states in the XXZ chain: Finite-size scaling probing symmetry-enriched $c=1$ conformal field theories
}
\author[1]{Chenan Wei}
\author[2,3]{Vagharsh V. Mkhitaryan}
\author[1]{Tigran A. Sedrakyan}
\affiliation[1]{Department of Physics, University of Massachusetts, Amherst, Massachusetts 01003, USA}
\affiliation[2]{University of Regensburg, Universitatsstrasse 31, 93053 Regensburg, Germany}
\affiliation[3]{Ames Laboratory, U.S. Department of Energy, Ames, Iowa 50011, USA}
\date{\today}
\abstract{We study the low-energy properties of the one-dimensional spin-1/2 XXZ chain with time-reversal symmetry-breaking pseudo-scalar chiral interaction and propose a phase diagram for the model. In the integrable case of the isotropic Heisenberg model with the chiral interaction, we employ the thermodynamic Bethe ansatz to find ``chiralization", the response of the ground state versus the strength of the pseudo-scalar chiral interaction of a chiral Heisenberg chain. Unlike the magnetization case, the chirality of the ground state remains zero until the transition point corresponding to critical coupling $\alpha_c=2J/\pi$ with $J$ being the antiferromagnetic spin-exchange interaction. The central-charge $c=1$ conformal field theories (CFTs) describe the two phases with zero and finite chirality. We show for this particular case and conjecture more generally for similar phase transitions that the difference between two emergent CFTs with identical central charges lies in the symmetry of their ground state (lightest weight) primary fields, \textit{i.e.}, the two phases are symmetry-enriched CFTs. At finite but small temperatures, the non-chiral Heisenberg phase acquires a finite chirality that scales with the temperature quadratically. We show that the finite-size effect around the transition point probes the transition.}

\begin{document}
\maketitle
\section{Introduction}


The one-dimensional (1D) spin-1/2 XXZ spin chain is a paradigmatic model in low-dimensional quantum condensed matter physics. The model can be solved exactly by algebraic Bethe ansatz (BA)\cite{takhtadzhan1979quantum,faddeev1984spectrum} and serves as a fundamental platform for exploring quantum spin dynamics and entanglement, as well as for studying quantum phase transitions, critical phenomena, and exotic states of matter in one dimension. 
 The eigenstates of the open XXZ chain under the specific constrained boundary condition (phantom roots criterion), namely the so called gapless phantom states, are also explored within the BA framework in XXZ chain\cite{zhang2021chiral,popkov2021phantom,zhang2021phantom}.
On the other hand, the spin dynamics in the model can be described by the generalized hydrodynamics\cite{castro2016emergent,bertini2016transport}.
The Heisenberg limit of the model manifests a dynamic exponent of $z=3/2$, which signals anomalous superdiffusion of the spin intermediate between diffusive and ballistic spin transport\cite{gopalakrishnan2019kinetic,ljubotina2019kardar,scheie2021detection}.
Other methods are also employed to understand out-of-equilibrium statistics in spin chains\cite{etienne2022outofequilibrium,senese2023outofequilibrium}.
The model possesses a line of critical phase in the continuum (thermodynamic) limit, realizing a continuous conformal field theory (CFT) with the central charge, $c=1$. In the isotropic Heisenberg case, this criticality can be easily seen as the $SU(2)$-symmetric spin coupling being the critical point between the easy plane phase and the easy axis (antiferromagnetic) phase. 
The underlying conformal symmetry dramatically simplifies the description of the critical phase of the model. It plays a crucial role in understanding the scaling behavior and correlations of the system as it approaches the critical phase. Besides, from the exact solution, the leading-order scaling for the ground state energy, $E \sim (1/4-\ln 2)N+O(1)$ and the excitation energy gap, $m \sim \frac{\pi^2}{2N}$, where $N\gg 1$ is the number of spins, has been established in the literature, see e.g., Refs.~\onlinecite{yang1966oneI,yang1966oneII,yang1966oneIII}. 
This scaling is also consistent with the CFT prediction \cite{cardy1986logarithmic,calabrese2009entanglement}. Numerical exact diagonalization also confirms the same result, even with a relatively small system size \cite{medeiros1991lanczos}. Notably, the exploration of the isotropic Heisenberg chain extends beyond the spin-1/2 system, encompassing analytical and numerical investigations \cite{babujian1983exact,affleck1989critical,fuhringer2008DMRG}.
Among many of them, Haldane postulated a conjecture: half-integer spin Heisenberg chains exhibit gapless behavior, while integer spin Heisenberg chains manifest an excitation gap \cite{haldane1983nonlinear}.
Many other spin chains, supporting both critical and gapped phases, along with their modifications are also extensively studied in the literature, see e.g., Refs.~\cite{michal2021exact,yang2022three,he2023integrable,hao2023exact,jana2023topological,shaikh2023phase,bravyi2015gapped,padmanabhan2024integrability}.

The XXZ spin chain model can be conveniently solved employing the Bethe ansatz\cite{yang1966oneI,yang1966oneII,yang1966oneIII,faddeev1996algebraic,korepin1997quantum,slavnov2019algebraic} technique, a powerful method of finding exact solutions for solvable quantum many-body systems, particularly in one spatial dimension. The original idea of Bethe ansatz, or the coordinate Bethe ansatz, is to assume a form of the wavefunction that consists of superpositions of plane waves with different momenta and impose the boundary conditions and the conservation laws to obtain a set of equations for the allowed momenta known as the Bethe equations.
These equations determine the energy spectrum and the eigenstates of the system. Bethe ansatz can also be used to calculate thermodynamic properties and correlation functions.
The transfer matrix method is another convenient way to construct the Bethe equations, and hence the energy spectrum and the eigenstates of the system.
With the transfer matrix method, the Hamiltonian of the Heisenberg chain is a conserved quantity that can be generated from the logarithmic derivative of the transfer matrix, or equivalently by the boost operator method\cite{leeuw2019classifying,leeuw2021yang}
$H = \frac{iJ}{2}\left.\frac{\mathrm{d}}{\mathrm{d}u}\ln T(u)\right|_{u=i/2}-\frac{JL}{4}$,
where $T(u)$ is the transfer matrix and $u$ is the spectral parameter
(the details of this construction are presented in \cref{dlnT}).

One of the important aspects of the antiferromagnetic Heisenberg chain is its magnetization, which reflects the alignment of spins in response to an external magnetic field.
The introduction of an external field perturbs the system, breaking the $SU(2)$ symmetry.
The external field triggers the spins to align parallel to the external field and develop a finite magnetization.
The magnetization of the chain, $m$, can be obtained through various methods. In specific situations, Bethe ansatz provides a computation avenue for calculating $m$\cite{takahashi1991correlation,granet2019analytical}  Additionally, applying the worm algorithm Monte Carlo offers another viable approach\cite{kashurnikov1999quantum}. Particularly, around the transition point, the magnetization exhibits a square root scaling\cite{kashurnikov1999quantum},
\begin{equation}
    m \sim
    \begin{cases}
        \frac12 - \frac{1}{\pi}\sqrt{h_c-h},\quad h<h_c, \\
        \frac12,\quad h\geq h_c,
    \end{cases}
\end{equation}
where $h$ is the external field coupled to the $z$ component of the spins, \textit{i.e.}, $h\sum_n{s}^z_n$, and $h_c$ is the critical field.
The imaginary magnetic field is also studied in the context of the Ising CFT\cite{xu2022ising,xu2023ising,lencses2023multicriticality}.

In this work, we address the question concerning the explicit breaking of the time-reversal symmetry in 1D $s=1/2$ chains by pseudo-scalar chirality. Consider the anisotropic spin-1/2 Heisenberg chain (the XXZ model) with pseudo-scalar chiral interaction, described by the Hamiltonian:
\begin{dmath} \label{XXZ}
    H_\Delta(J,\alpha) =
    J \sum_{n} ({s}^x_n {s}^x_{n+1}+{s}^y_n {s}^y_{n+1}+\Delta {s}^z_n {s}^z_{n+1})
    + \alpha \sum_{n} \mathbf{s}_n \cdot (\mathbf{s}_{n+1} \cross \mathbf{s}_{n+2}),
\end{dmath}
where the spin-1/2 operators are given in terms of Pauli matrices as $s^\mu=\frac{\hbar}{2}\sigma_\mu$, $\mu=x,y,z$.
The Hamiltonian with parameters $\alpha$ and $-\alpha$ can be mapped to each other with all the spins flipped. Hence, without loss of generality, we consider $\alpha$ positive throughout the paper.
Importantly, in the isotropic Heisenberg limit, $\Delta\to 1$ (XXX model), the pseudo-scalar chiral interaction can arise from the $t/U$ expansion of the half-filled Hubbard model\cite{takahashi1977half,macDonald1988expansion,motrunich2006orbital}.
Furthermore, in this limit, the total chirality, $\mathcal{C} = \sum_n \mathbf{s}_{n} \cdot (\mathbf{s}_{n+1} \cross \mathbf{s}_{n+2})$, is a conserved quantity which can be generated by the transfer matrix:
     $\mathcal{C} = \left. -\frac{i}{4} \frac{d^2}{du^2} \ln T(u) \right|_{u=i/2}
     +\frac{1}{4} iL$
(see \cref{dlnT} for the derivation). Therefore, the model \cref{XXZ} is exactly solvable in the isotropic Heisenberg limit. In the following, we derive analytical results for the isotropic Heisenberg limit and investigate the more general, anisotropic XXZ case numerically.

Another interesting aspect of the chirality transition in the Heisenberg chain with pseudo-scalar chiral interaction, adding to the motivation for the present study, is its consideration from the standpoint of commensurate-incommensurate phase transition  demonstrated by Tsvelik in Ref.~\onlinecite{tsvelik1990incommensurate}. According to this picture, the phase with a non-zero chirality, corresponding to a symmetry-enriched, $c=1$ CFT, possesses an incommensurate ground state within the spin chain. This hints at a connection of incommensurate phase transitions to symmetry-enriched CFTs, calling for further scrutiny. 

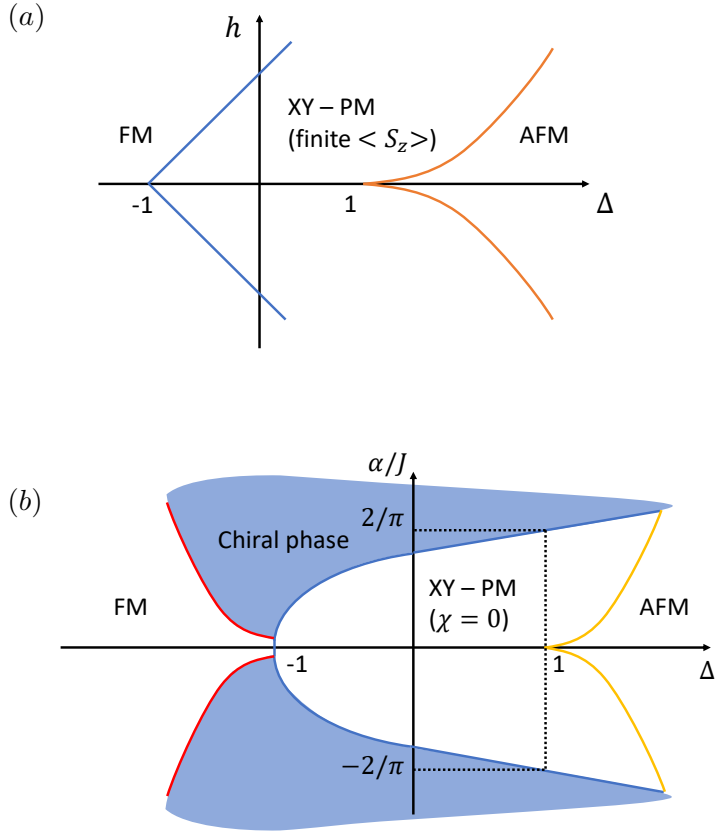
\begin{figure}
    \centering
    \begin{minipage}{0.69\textwidth}
    \includegraphics[width=\linewidth]{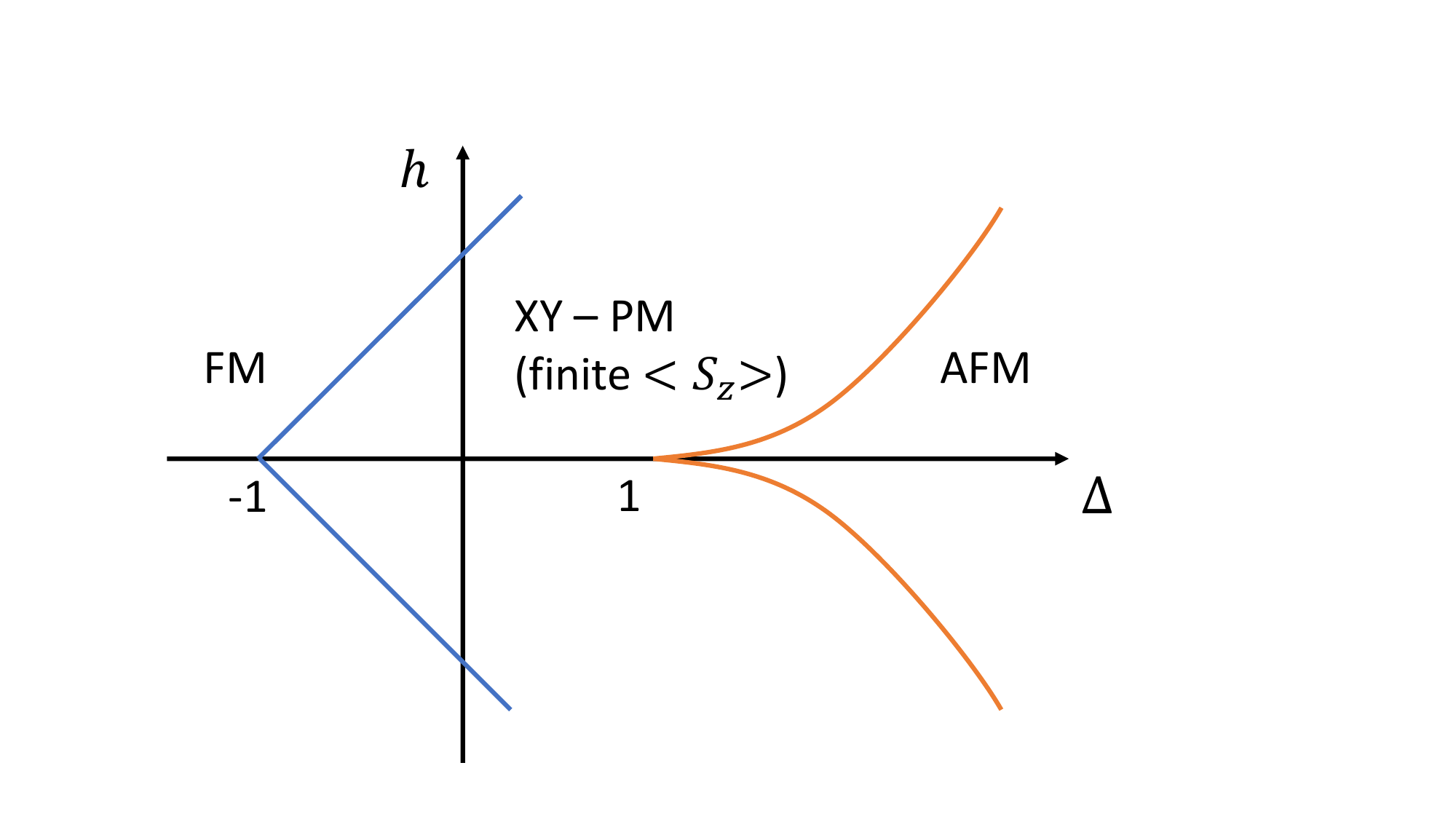}
    \begin{picture}(0,0)
    {\def\unitlength{}
    \ifTwocolumn
        \put(-0.5\linewidth,0.5\linewidth){$(a)$}
    \else
        \put(0\linewidth,0.5\linewidth){$(a)$}
    \fi
    }
    \end{picture}
    \phantomsubcaption{\label{phase_diagrams_Heisenberg}}
    \end{minipage}
    \begin{minipage}{0.69\textwidth}
    \includegraphics[width=\linewidth]{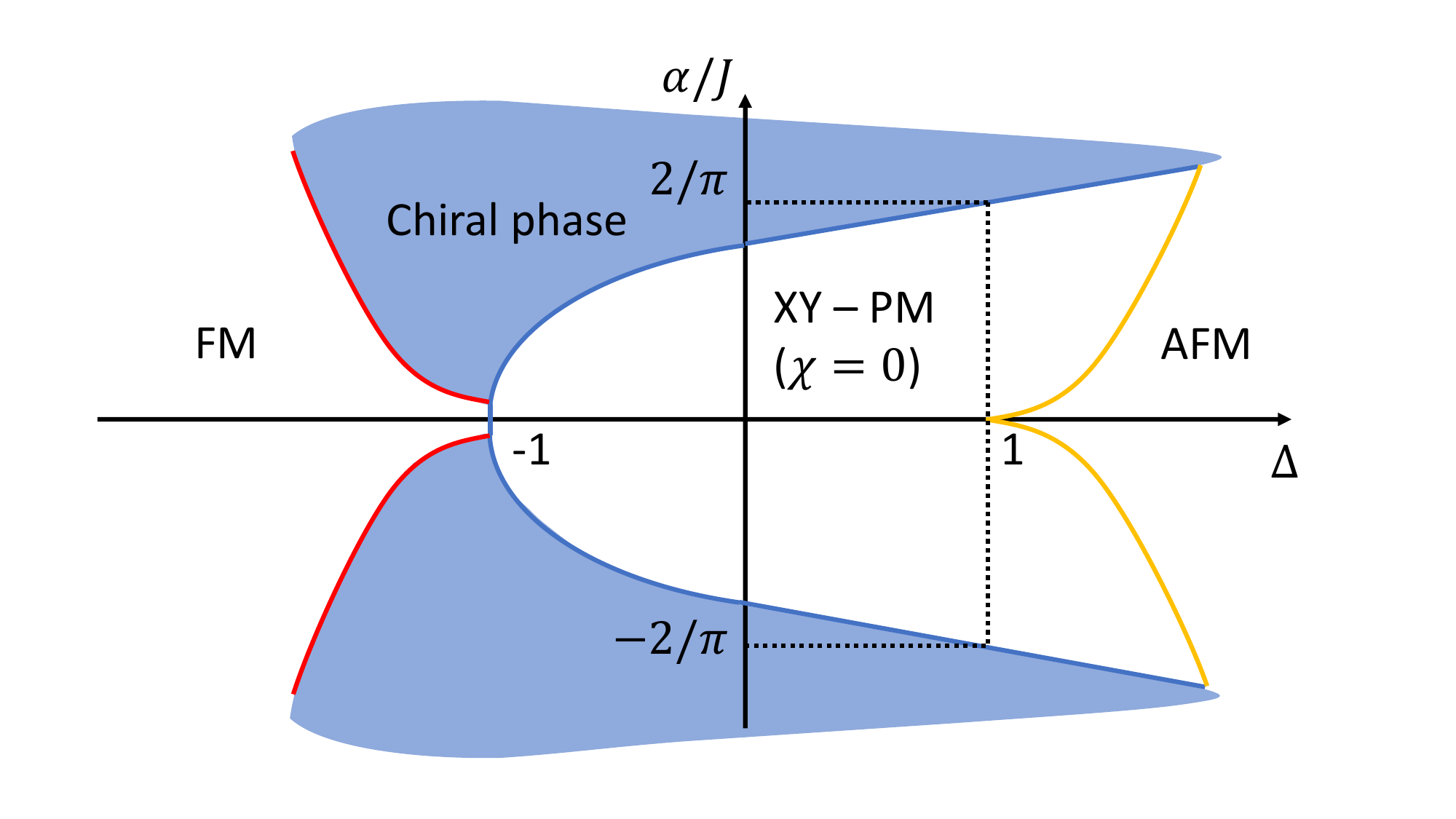}
    \begin{picture}(0,0)
    {\def\unitlength{}
        \ifTwocolumn
        \put(-0.5\linewidth,0.5\linewidth){$(b)$}
    \else
        \put(0\linewidth,0.5\linewidth){$(b)$}
    \fi
    }
    \end{picture}
    \phantomsubcaption{\label{phase_diagrams_chiral}}
    \end{minipage}
    \caption{\label{phaseDiagram}
    Schematic zero temperature phase diagrams of time-reversal symmetry broken XXZ spin chains. Panel \protect\subref{phase_diagrams_Heisenberg} shows the phase diagram of an anisotropic Heisenberg chain with the external field. Panel \protect\subref{phase_diagrams_chiral} shows the phase diagram of the chiral chain, with FM denoting the Ising ferromagnetic phase, XY - PM the XY paramagnetic phase, and AFM -- the Ising antiferromagnetic phase. The dotted vertical line corresponds to the exactly solvable Heisenberg model with scalar chirality term.}
\end{figure}

Given the property of conservation of the total chiralization in XXX model, resembling the similar property of the magnetization in the XXZ model, one might naturally anticipate that the pseudo-scalar chiral interaction leads to a phase diagram of the Heisenberg model similar to that in the case of a magnetic field. However, here we derive a different scenario where the time reversal $\mathcal{T}$ and parity $\mathcal{P}$ are both broken by the scalar 
pseudo-scalar chiral interaction, whereas the combination of the two symmetries, $\mathcal{P} \mathcal{T}$, remains preserved. We show that the chiralization exhibits a critical behavior different from the magnetization on both sides of the transition point. Specifically, in the integrable isotropic Heisenberg limit, at the transition to the chiral phase, the chirality $\chi= \langle\mathbf{s}_n \cdot (\mathbf{s}_{n+1} \cross \mathbf{s}_{n+2}) \rangle$ scales linearly with the strength of the pseudo-scalar chiral interaction $\alpha$:
\begin{equation}
\label{chiscale}
    \chi \simeq
    \begin{cases}
        0,\quad \alpha<\alpha_c, \\
        -0.74(\alpha - \alpha_c),\quad \alpha\gtrsim\alpha_c.
    \end{cases}
\end{equation}
Thus the chirality remains zero while $\alpha$ does not exceed the critical threshold $\alpha_c=2J/\pi$ and increases linearly beyond the transition point. Furthermore, both phases with zero and non-zero chirality are gapless and can be described by conformal field theories with central charge $c=1$. We further generalize the described result to the case of the XXZ chain, leading to the phase diagram depicted in \cref{phaseDiagram}.

The possibility of strong pseudo-scalar chiral interaction simply suppressing one of the chiral copies of the corresponding CFT and leading to a chiral CFT is ruled out because the chiral central charge, ($c_L - c_R$), of a CFT in 1+1 dimensions corresponding to a lattice model is shown to be zero \cite{kapustin2019absence,zou2022modular,fan2022from,hellerman2021quantum,ji2022unified,wiegmann2022chiral}.
Hence, we conclude that the crux of the distinction between the two phases with and without chiral order lies in their respective symmetries. The combination of the conformal symmetry with some additional symmetries is known to produce a multitude of distinct CFTs. For instance, a $U(1)$ symmetry leads to an extended CFT featuring an emergent Kac-Moody algebra \cite{wang2022emergence}. In addition, one can create a novel CFT by modding out a part of the symmetry, a concept known as orbifold CFT\cite{ginsparg1988curiosities,dijkgraaf1989operator,chen2017from}.

The chiralization transition considered in this paper
can be interpreted as a transition between symmetry-enriched CFTs.
Symmetry-enriched CFTs combine the concept of CFT with symmetries, particularly focusing on how global symmetries can enrich the structure and properties. Symmetry enrichments in CFTs can lead to new types of quantum critical behavior and provide insights into the classification of gapless topological phases\cite{verresen2021gapless,ke2022universal,yu2022conformal}.
To outline the transition, note that 
for small values of $\alpha$ ($\alpha < \alpha_c$), the primary field associated with the emergent CFT is related to the $x$-component of the spin operator $s_x$ (spin flip). This operator flips the sign under the time reversal,
\begin{equation}
(\mathcal{PT})^{-1} s^x \mathcal{PT} = -s^x.
\end{equation}
Conversely, for larger values of $\alpha$ ($\alpha \gg \alpha_c$), the ground state primary field is related to the phase twist $p$ discussed in detail in \cref{EDchiral}:
\begin{equation} \label{phasetwist}
p \equiv \prod_{n=1}^L P_n\left(\frac{2\pi n}{N}-\pi \Pi\left[n(N + 1)\right]\right)
\end{equation}
where $P_n(\varphi)$ is the phase shift gate on site $n$ that maps the basis states $\ket{\downarrow} \to \ket{\downarrow}$ and $\ket{\uparrow} \to e^{i\varphi}\ket{\uparrow}$. The function $\Pi: \mathbb{Z}\rightarrow\{0,1\}$ is the parity of the argument.
In explicit matrix form,
\begin{dmath}
    P_n(\varphi) =
        \begin{pmatrix}
            1 & 0\\
            0 & e^{i\varphi}
        \end{pmatrix}.
\end{dmath}
The phase shift is different at different sites, making the phase of the excitation a twisting pattern.
The phase twist $p$ remains invariant under time reversal:
\begin{equation}
(\mathcal{PT})^{-1} p (\mathcal{PT}) = p.
\end{equation}

The different behavior of low-energy excitations under $\mathcal{PT}$ symmetry makes it impossible to smoothly connect these two Heisenberg-like and chiral phases, giving rise to a transition between them. This should be contrasted with the conventional criticality characterized by the interplay between symmetry-breaking and symmetry-preserving mechanisms.
Introducing a simple solvable interacting model, which is experimentally relevant and encompasses transitions described by symmetry-enriched CFT, would significantly advance the translation of these theoretical concepts into practical, real-world applications.

In this paper, we show that the transition between the two phases can be detected via the universal finite-size scaling behavior of the excitation gap, the ground state energy, and the entanglement entropy. The theoretical underpinning of finite-size scaling in time-reversal symmetric CFTs, supported by numerical studies, is well-established\cite{affleck1986universal,gulden2016universal}. 
The universal finite size effects in time-reversal-symmetry broken CFTs have recently been derived for $c=1/2$ theories. In particular, it was shown that the terms in the excitation gap and the ground state energy relations, inversely scaling with the system size, appear universal and depend only on the central charge of the Virasoro algebra, $c$, and scale with the time-reversal symmetry-breaking term universally. For the ground state energy, $E$, of $N$ non-interacting particles with open or periodic boundary conditions, one obtains\cite{wang2020universal}
\begin{equation}
\label{Egni}
    E / J = \epsilon N + b + \frac{cv}{N} \left[g(N m, \zeta h^2 N) - \theta \right] + O\left(\frac{1}{N\ln N}\right),
\end{equation}
where $\epsilon$ is the bulk energy per spin, $b$ is the boundary energy, $c$ is the central charge of the corresponding CFT, $v$ is the effective velocity, $m$ is the spectral gap (the critical CFT behavior sets in in the $m=0$ limit), 
$\zeta$ is a model-dependent function of the localization length $\xi$, $h$ is the external time-reversal symmetry breaking field (can be the magnetic field), and $g(x)$ is a model-independent universal function. The parameter $\theta$ depends on the boundary conditions: $\theta=\frac{\pi}{24}$ for the system with open boundary conditions, and $\theta = \frac{\pi}{6}$ for the system with periodic boundary conditions.


We numerically demonstrate that in the interacting case of the Heisenberg model with pseudo-scalar chiral interaction, \cref{XXZ} with $\Delta=1$, the critical $(m=0)$ scaling in \cref{Egni}
preserves its form with the following time-reversal-symmetry breaking field:
\begin{equation}
\label{Egint}
h\rightarrow \left\{
\begin{matrix}
\sqrt{\alpha-\alpha_c},& \text{for} \;\; \alpha\geq \alpha_c\\
0,  & \text{otherwise}.
\end{matrix}
\right.
\end{equation}
Thus, the scaling variable in the argument of the universal scaling function $g$ of \cref{Egni} becomes $\zeta h^2 N \rightarrow \zeta (\alpha-\alpha_c) N$, with other terms preserving their physical meanings.
\Cref{Egint} is one of the main results of this paper, with function $g$ playing the role of the Landau order parameter in describing the quantum phase transition.

Although \cref{XXZ} is integrable when $\Delta = 1$, for general $\Delta$, it is no longer integrable.
Based on our analytical and numerical calculations, we draw the phase diagram of the anisotropic Heisenberg spin-1/2 chain (XXZ model) in the presence of the pseudo-scalar chiral interaction and 
compare it to the well-known magnetization phase diagram of the anisotropic Heisenberg chain in \cref{phaseDiagram}.
Both the pseudo-scalar chiral interaction and the coupling to external magnetic field break the time-reversal symmetry at the level of the Hamiltonian. \Cref{phaseDiagram} illustrates that once the chiral term replaces the external magnetic field, the whole phase diagram of the model changes drastically.
Particularly, time-reversal symmetry-breaking chiral phases take over from both the ferromagnetic and antiferromagnetic Heisenberg domains, with the corresponding tricritical points. At the same time, the underlying state remains similar to that of the isotropic Heisenberg critical phase with small $\alpha<\alpha_c$.

The remainder of the paper is organized as follows. In \cref{chiral_spin_chian}, we outline the basic properties of the model with the chirality term and describe the derivation of the scalar chirality order parameter, the bosonization of the model, and the qualitative description of the chiralization phase transition for the isotropic Heisenberg case. 
In \cref{BA_solution}, we present the details of the algebraic and thermodynamic Bethe ansatz solution of the isotropic Heisenberg model with the pseudo-scalar chiral interaction, derive the exact expression for the chirality order parameter, and discuss the quantum phase transition. At finite temperatures, we find that the chirality scales with the temperature quadratically for all values of $\alpha$, irrespective of the ground state chirality. In 
\cref{XXZsect}, we discuss our numerical data obtained for the XXZ model with the pseudo-scalar chiral interaction, and in 
\cref{finite_size_scaling}, we present the DMRG calculations of the finite size scaling of various characteristics of the XXZ model. Section \ref{conclusion} contains our concluding remarks and outlooks. Details of analytical calculations are presented in the appendices.

\section{The Heisenberg chain with chirality} \label{chiral_spin_chian}

We start with the $SU(2)$ symmetric case of the Hamiltonian \cref{XXZ}, namely a one-dimensional spin-1/2 Heisenberg chain with the pseudo-scalar chiral interaction, described by the Hamiltonian,
\begin{equation} \label{hamiltonian}
\begin{aligned}
    H_1(J, \alpha) =
    J\sum_{n} \mathbf{s}_n \mathbf{s}_{n+1}
    + \alpha \sum_{n} \mathbf{s}_n \cdot (\mathbf{s}_{n+1} \cross \mathbf{s}_{n+2}),
\end{aligned}
\end{equation}
where the periodic boundary conditions are imposed.
Importantly, the chiral term in \cref{hamiltonian} commutes with the first Heisenberg exchange term, and hence, the model can be solved and the free energy evaluated within the same thermodynamic Bethe Ansatz (TBA) framework that provides the solution of the exchange part only \cite{tsvelik1990incommensurate,mkhitaryan2006mean, mkhitaryan2008next} (for a pedagogical review of TBA, see, e.g., Refs. \cite{tsvelik1990incommensurate,takahashi1997thermodynamical}). In the phase diagram \cref{phaseDiagram}(b), the integrable region includes the vertical line at $\Delta=1$, the horizontal axis with $\alpha=0$, and the line $J \rightarrow 0$. 
The model \cref{hamiltonian} can be generalized to the anisotropic case, and an exactly solvable model can be formulated by defining a conserved anisotropic chirality for the XXZ Hamiltonian preserving the integrability. This generalization is discussed in Ref.~\cite{tsvelik1990incommensurate} in detail. Note, however, that it is different from \cref{XXZ}, in which the last chiral term is not conserved.
The same can be applied to various quantum integrable chains termed staggered integrable ladder models\cite{sedrakyan2001staggered,ambjorn2001integrable,arnaudon2001generalization}. 
Exploring the stabilization of quantum phases with finite anisotropic chirality would also be useful as it could help to understand the nature of the quantum chiral spin liquid in two-dimensional frustrated systems\cite{sedrakyan2009fermionic,sedrakyan2012composite,sedrakyan2014absence,sedrakyan2015statistical,sedrakyan2015spontaneous,wang2018chern,maiti2019fermionization,wang2022emergent,wei2023chiral,wang2023excitonic}.

The exact evaluation of the chiral order parameter in a model \cref{hamiltonian} can be achieved by noting that once the free energy per spin, $F(J,\alpha)$, is evaluated, the chirality per spin can be found using the Hellman-Feynman theorem\cite{feynman1939forces}, 
by taking the derivative of the free energy,
\begin{equation} \label{chi}
\begin{aligned}
    \chi= \langle\mathbf{s}_n \cdot (\mathbf{s}_{n+1} \cross \mathbf{s}_{n+2}) \rangle =\partial_\alpha F(J, \alpha).
\end{aligned}
\end{equation}
Before we present the TBA solution of the model \cref{hamiltonian}, it is instructive to discuss the bosonization of the Hamiltonian, illustrating the CFT nature of the model. For the details of this technique, we refer to Ref.~\onlinecite{giamarchi2003quantum}.

The bosonization of the Heisenberg (exchange) part of \cref{hamiltonian} is standard and is briefly discussed in \cref{BRG}. Applying the Jordan-Wigner transformation and taking the continuous limit to the leading order in bosonic fluctuations, one gets
\begin{equation}
\begin{aligned}
    H_1(J, 0) &\sim \frac 12 v \int dx [ g^{-1} (\partial_x \phi)^2 + g (\partial_x \theta)^2],
\end{aligned}
\end{equation}
where $v$ is the effective velocity, $g$ is the Luttinger parameter, and $\phi$ and $\theta$ are the boson and dual boson fields. Along the same lines, 
one can bosonize the chiral term, yielding
\begin{equation}
\begin{aligned}
    H_{\chi} \equiv H_1(0, \alpha) &\sim \frac{1}{2} \alpha \int dx \partial_x \phi \partial_x \theta,
\end{aligned}
\end{equation}
where we have dropped the irrelevant terms.
The subsequent perturbative renormalization group analysis\cite{giamarchi2003quantum} shows that the chiral term is marginal to the Heisenberg part so that, depending on the magnitude of $\alpha$, the effective CFT describing the model is either the same as that for $H_1(J, 0)$ or a different one, which captures the non-zero chiral ordering.

A similar bosonization analysis for conserved quantities generated from higher order derivative of the transfer matrix, such as the four-spin term, shows they are irrelevant. So, the ground state will remain the Heisenberg criticality upon introducing such perturbations.
The same can be seen in the classical limit.
Suppose the classical spins of magnitude $s$ are arranged so the angles between nearest-neighbor and next nearest-neighbor spins are $\pi - \theta$ and $\theta$ separately (fluctuation above the antiferromagnetic phase). The energy increase due to the exchange interaction is $\delta E_J = J s^2 (\cos(\pi-\theta) - \cos(\pi)) \approx \frac12 Js^2 \theta^2$. And the energy decrease because of the pseudo-scalar chiral interaction is
$
\delta E_{\alpha}
= -\alpha s^3
\sqrt{1+2\cos\theta \cos^2(\pi-\theta)
- \cos ^{2}\theta
- 2\cos^{2}(\pi-\theta)}
\approx
- \frac{\sqrt{3}}{2} \alpha s^3 \theta^2.
$
For a large $\alpha > J/\sqrt3 s$, the total energy change $\delta E = (\frac12 Js^2 - \frac{\sqrt{3}}{2} \alpha s^3) \theta^2$ is negative. So, the chiral state is favored. This estimate suggests that classically, $\alpha_c=(J/s)/\sqrt3 \simeq 0.57735 (J/s)$, which is off by $O(10)^{-1}$ from the quantum value $\alpha_c=(2/\pi)J$, to be derived in the next section, after rescaling the exchange coupling $J$ by large $s$. When higher-order conserved quantities are introduced, the change of total energy is dominated by the increase in exchange interaction, leading to an antiferromagnetic phase.

\section{The Bethe ansatz solution of the Heisenberg model with chirality} \label{BA_solution}

In this section, we closely follow the notations used in the existing literature on TBA \cite{mkhitaryan2006mean, takahashi1997thermodynamical}, and defer derivation details to \cref{TBA}. The thermodynamic Bethe equations for the chiral Heisenberg chain form a set of non-linear integral equations on the so-called $n$-string quasienergies, $\epsilon_n(x)$, parameterized by integer $n=1, 2,\cdots$, and real $x$:
\ifTwocolumn
    \begin{equation} \label{epsilon}
    \begin{aligned}
        &\epsilon_1(x) = -2 \pi (J - \alpha \partial_x) s(x)+ T s * \ln \left(1+\exp \left(\frac{\epsilon_2(x)}{T}\right)\right), \\
        &\epsilon_n(x) = \\
        &T s * \ln \left[ \left(1+\exp \left(\frac{\epsilon_{n-1}(x)}{T}\right)\right) \left(1+\exp \left(\frac{\epsilon_{n+1}(x)}{T}\right)\right) \right],\quad n\geq 2,\\
        &\lim _{n \rightarrow \infty} \frac{\epsilon_n(x)}{n} = 0,
    \end{aligned}
    \end{equation}
\else
    \begin{dmath} \label{epsilon}
        {\epsilon_1(x) = -2 \pi (J - \alpha \partial_x) s(x)+ T s * \ln \left(1+\exp \left(\frac{\epsilon_2(x)}{T}\right)\right)}, \\
        {\epsilon_n(x) = T s * \ln \left[ \left(1+\exp \left(\frac{\epsilon_{n-1}(x)}{T}\right)\right) \left(1+\exp \left(\frac{\epsilon_{n+1}(x)}{T}\right)\right) \right]},\quad n\geq 2,\\
        {\lim _{n \rightarrow \infty} \frac{\epsilon_n(x)}{n} = 0},
    \end{dmath}
\fi
where $T$ is the temperature (the Boltzmann constant is set to 1), $s(x)=\frac{1}{4}\sech(\frac\pi 2 x)$, and $``*"$ denotes the convolution, $f*g(x)\equiv\int_{-\infty}^{\infty}f(x-y)g(y)dy$.
The free energy per spin is given by
\ifTwocolumn
    \begin{equation} \label{FreeenergyT}
    \begin{aligned}
        F =&-J\left(\ln2-\frac{1}{4}\right) \\
        &-T\int_{-\infty}^{\infty} s(x)\ln \left(1+\exp \left(\frac{\epsilon_1(x)}{T}\right)\right)dx.
    \end{aligned}
    \end{equation}
\else
    \begin{dmath} \label{FreeenergyT}
        F =-J\left(\ln2-\frac{1}{4}\right)
        -T\int_{-\infty}^{\infty} s(x)\ln \left(1+\exp \left(\frac{\epsilon_1(x)}{T}\right)\right)dx.
    \end{dmath}
\fi
\Cref{epsilon,FreeenergyT} determine the free energy, and hence, other physical properties of the chiral Heisenberg chain. In the following, we find the exact solution of \cref{epsilon,FreeenergyT} in the zero-temperature limit, $T \rightarrow 0$, and deduce the low-temperature free energy perturbatively.

\subsection{The ground state chirality}

The ground state is found in the zero temperature limit,
$T \to 0$. In this limit, the Bethe equations and the free energy are simplified to
\begin{equation} \label{T0epsilon}
\begin{aligned}
&\epsilon_{1}(x)=-2\pi (J - \alpha \partial_x) s(x)+s*\epsilon_{2}^{+}(x), \\
&\epsilon_{n}(x)=s*(\epsilon_{n-1}^{+}(x)+\epsilon_{n+1}^{+}(x)), \quad n\geq 2, \\
&\lim_{n\to\infty}\frac{\epsilon_n(x)}{n}=0,
\end{aligned}
\end{equation}
and
\begin{equation} \label{freeenergy}
F=-J\left(\ln 2-\frac{1}{4}\right)-\int_{-\infty}^{\infty} s(x) \epsilon_1^{+}(x) d x,
\end{equation}
where
\begin{equation}
\epsilon_n^{+}(x) \equiv \begin{cases}\epsilon_n(x), & \text { for } \epsilon_n(x) \geq 0, \\
0, & \text { for } \epsilon_n(x)<0. \end{cases}
\end{equation}
Thus, in \cref{T0epsilon}, only $\epsilon_1$ can take negative values. 

Depending on the ratio, $\alpha / J$, the free energy takes different forms.  For $\alpha < \alpha_c = 2J/\pi$, the system \cref{T0epsilon} yields the solution with $\epsilon_1<0$ and $\epsilon_n=0$, $n\geq 2$. Therefore, the second term in \cref{freeenergy} disappears and $f$ is independent on $\alpha$:
\begin{equation}
    F(J, \alpha)=-J\left(\ln 2 - \frac14\right).
\end{equation}
Hence, according to \cref{chi}, the chirality is zero:
\begin{equation}
    \chi(\alpha)=0,\quad \alpha < \alpha_c = 2J/\pi.
\end{equation}

For $\alpha > \alpha_c = 2J/\pi$, on the other hand, $\epsilon_1(x)$ is monotonically increasing and undergoes a sign change. In this case, \Cref{T0epsilon} is solved by the Wiener-Hopf method \cite{tsvelik1990incommensurate, mkhitaryan2006mean}. Suppose $\epsilon_1(x)$ crosses the zero at $x=a$, \textit{i.e.}, $\epsilon_1(a)=0$.
Then, the symmetry-breaking ground state energy can be written as
\ifTwocolumn
    \begin{equation}\label{falpha}\begin{aligned}
        F(\alpha) / J &=-(\ln 2 - \frac14) + \frac12\sum_{n,k=0}^\infty(-1)^{k+n}[1-\frac{\pi \alpha}{J} (k+1/2)]\\&\times\frac{e^{-\pi a(k+n+1)}G(n)G(k)}{k+n+1},
    \end{aligned}\end{equation}
\else
    \begin{dmath}\label{falpha}
        F(\alpha) / J =-\left(\ln 2 - \frac14\right) + \frac12\sum_{n,k=0}^\infty(-1)^{k+n}\left[1-\frac{\pi \alpha}{J} (k+1/2)\right]
        \frac{e^{-\pi a(k+n+1)}G(n)G(k)}{k+n+1},
    \end{dmath}
\fi
where
\begin{equation}
    G(k)=\sqrt{2\pi}
    \frac
    {\exp\{-k-1/2+(k+1/2)\ln(k+1/2)\}}
    {\Gamma(k+1)}
\end{equation}
is defined through the gamma function $\Gamma$,
and $a$ satisfies
\begin{equation} \label{aofalpha}
    \sum_{k=0}^\infty(-1)^k\left[1-\frac{\pi\alpha}{J}(k+1/2)\right]e^{-\pi a (k+1/2)}G(k)=0,
\end{equation}
or equivalently,
   \begin{equation} \label{alphaa}
   \begin{aligned}
       \alpha / J = \frac{\sum_{k=0}^\infty(-1)^k e^{-\pi a (k+1/2)}G(k)}{\sum_{k=0}^\infty(-1)^k[\pi (k+1/2)]e^{-\pi a (k+1/2)}G(k)}.
   \end{aligned}
   \end{equation}
For the chirality per spin, we find the exact expression,
\begin{equation} \label{chiSeries}
\begin{aligned}
    \chi &= -\frac\pi4 \Bigl( \sum_{n}(-1)^{n} e^{-\pi a(n+1/2)} G(n) \Bigr )^2,
\end{aligned}
\end{equation}
using \cref{chi} (see \cref{chiralityLargeAlpha}).

Thus, we conclude that the ground state chirality is zero for small $\alpha$ up to $\alpha_c = 2J/\pi$, where the chiralization transition occurs, leading to a non-zero ground state chirality for $\alpha>\alpha_c$. Upon numerically solving the transcendental equations \cref{alphaa,aofalpha} for $a$, we find that in the vicinity of the quantum phase transition, the analytical expression for chirality acquires a linear asymptote:
\begin{equation} \label{chasym}
\chi = - 
\zeta_0\; (\alpha/J - 2/\pi),
\end{equation}
where the linearity coefficient is $\zeta_0\approx 0.740308$, producing the asymptotic behavior \cref{chiscale}. 
In \cref{closedform}, the exact closed form expression for $\chi$ is found, and the derivation details of the scaling of $\chi$ around the critical point, given by \cref{chasym}, are presented.


\subsection{Finite temperature behavior}

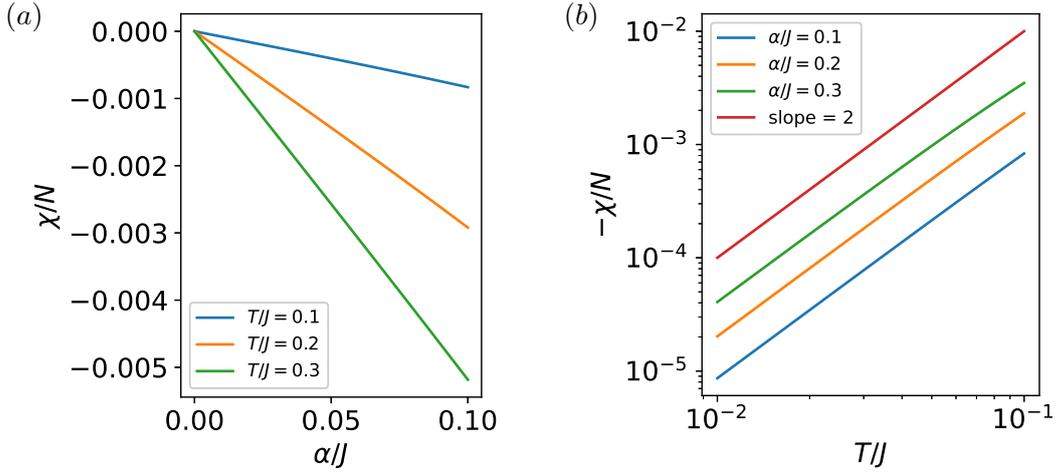
\begin{figure}
    \centering
    \hfill
    \begin{minipage}{0.45\textwidth}
    \includegraphics[width=\linewidth]{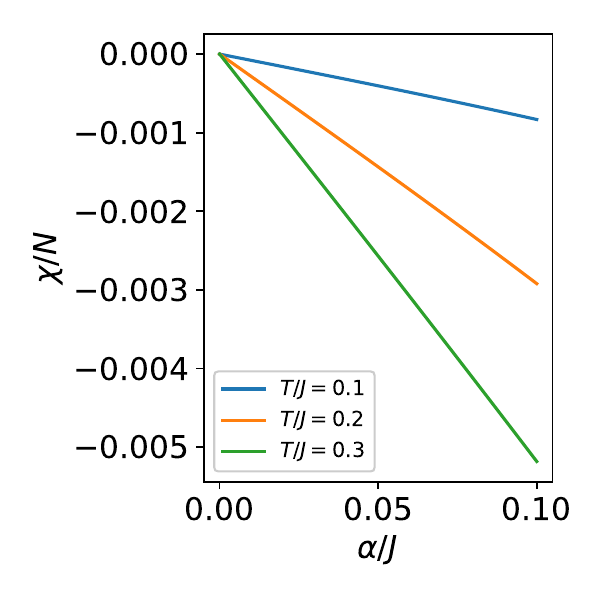}
    \begin{picture}(0,0)
    {\def\unitlength{}
    \ifTwocolumn
        \put(-0.5\linewidth,1.0\linewidth){$(a)$}
    \else
        \put(0\linewidth,1.0\linewidth){$(a)$}
    \fi
    }
    \end{picture}
    \phantomsubcaption{\label{chialphaT}}
    \end{minipage}
    \hfill
    \begin{minipage}{0.45\textwidth}
    \includegraphics[width=\linewidth]{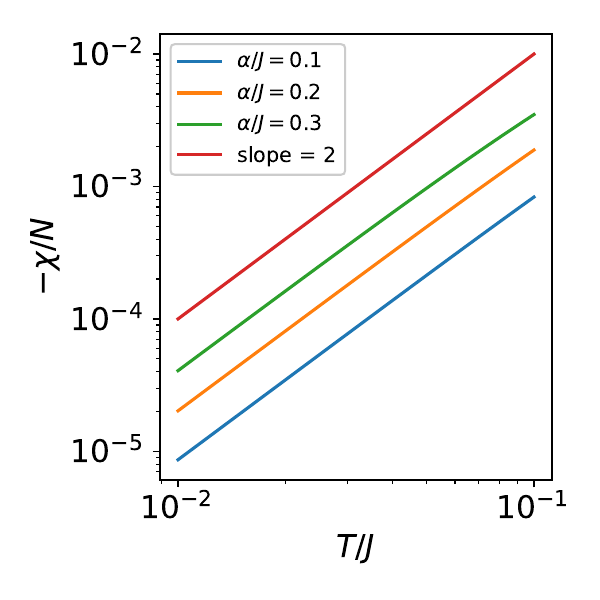}
    \begin{picture}(0,0)
    {\def\unitlength{}
    \ifTwocolumn
        \put(-0.5\linewidth,1.0\linewidth){$(b)$}
    \else
        \put(0\linewidth,1.0\linewidth){$(b)$}
    \fi
    }
    \end{picture}
    \phantomsubcaption{\label{chiT}}
    \end{minipage}
    \hfill
    \begin{minipage}{0\textwidth}
    \end{minipage}
    \caption{\label{finiteT} 
Finite temperature behavior of the chirality per site at low temperatures and small $\alpha$ computed from \cref{chiTIntegral}. Panel \protect\subref{chialphaT} shows the linear dependence of $\chi$ on $\alpha$, for three different temperatures, and panel \protect\subref{chiT} is plotted in log-log scale, with slopes showing the quadratic dependence of chirality on the temperature, for four different values of $\alpha$.}
\end{figure}

Before we proceed further and explore properties of the ground states of the chiral Heisenberg model, we look at the finite temperature corrections in the small- $\alpha$ (non-chiral) phase.
To study the behavior at low temperatures ($\Delta_T(\epsilon_i) \ll \epsilon_i$ with $\Delta_T(\epsilon_i)$ specified below), thermodynamic BE equations \ref{epsilon} are rewritten as
\ifTwocolumn
    \begin{equation}
    \begin{aligned}
    \epsilon_{1}(x) &= -2\pi (J - \alpha \partial_x) s(x)+s*\epsilon_{2}^{+}(x) + s*\Delta_T(\epsilon_2), \\
    \epsilon_{n}(x) &= s*(\epsilon_{n-1}^{+}(x)+\epsilon_{n+1}^{+}(x)) \\
    &+ s*(\Delta_T(\epsilon_{n-1}) + \Delta_T(\epsilon_{n+1})), \\
    &\lim_{n\to\infty}\frac{\epsilon_n(x)}{n}=0,
    \end{aligned}
    \end{equation}
\else
    \begin{equation}
    \begin{aligned}
    \epsilon_{1}(x) &= -2\pi (J - \alpha \partial_x) s(x)+s*\epsilon_{2}^{+}(x) + s*\Delta_T(\epsilon_2), \\
    \epsilon_{n}(x) &= s*(\epsilon_{n-1}^{+}(x)+\epsilon_{n+1}^{+}(x))
    + s*(\Delta_T(\epsilon_{n-1}) + \Delta_T(\epsilon_{n+1})), \\
    &\lim_{n\to\infty}\frac{\epsilon_n(x)}{n}=0,
    \end{aligned}
    \end{equation}
\fi
where
\begin{equation}
    \Delta_T(x) =
    \begin{cases}
        T \ln [1+\exp(x/T)], \quad x<0, \\
        T \ln [1+\exp(x/T)] - x, \quad x\geq 0,
    \end{cases}
\end{equation}
and the free energy is rewritten as
\begin{dmath}
    f=-J\left(\ln 2-\frac{1}{4}\right) - \int_{-\infty}^{\infty} s(x) \epsilon_1^{+}(x) d x
    -\int_{-\infty}^{\infty} s(x) \Delta_T(\epsilon_1) dx.
\end{dmath}
Now, one can treat the terms with $\Delta_T(\epsilon_i)$ perturbatively, 
\begin{equation}
\epsilon_{1} = \epsilon_{1}^{(T=0)} + s*\Delta_T(\epsilon_2^{(T=0)}) = \epsilon_{1}^{(T=0)} + s*\Delta_T(a_{1}*\epsilon_1^{+(T=0)}),
\end{equation}
yielding the free energy,
\begin{dmath}\label{fT0}
    f = f^{(T=0)} - \int_{-\infty}^{\infty} s^2 * \Delta_T(a_{1}*\epsilon_1^{+(T=0)}) d x 
    -\int_{-\infty}^{\infty} s(x) \Delta_T(\epsilon_{1}^{(T=0)}) dx.
\end{dmath}
For $\alpha < 2J/\pi$, the zero-temperature quasienergy $\epsilon_1^{(T=0)} = -2\pi (J - \alpha \partial_x) s(x)$ is negative, so the second term of \cref{fT0} vanishes, giving
\begin{equation}
\begin{aligned}
    f = f^{(T=0)} -\int s(x) \Delta_T(\epsilon_{1}^{(T=0)}) dx.
\end{aligned}
\end{equation}
Since $f^{(T=0)}$ is a constant independent on $\alpha$, we get:
\begin{equation} \label{chiTIntegral}
\begin{aligned}
    \chi/N = \frac{\partial f}{\partial \alpha} &= -\int_{-\infty}^{\infty} s(x) \Delta_T'(\epsilon_{1}^{(T=0)}) \frac{d \epsilon_{1}^{(T=0)}}{d \alpha} dx \\
    &= -2\pi \int_{-\infty}^{\infty} s(x) \Delta_T'(\epsilon_{1}^{(T=0)}) s'(x) dx.
\end{aligned}
\end{equation}
A numerical integral of \cref{chiTIntegral} gives us the low temperature $T$ and small $\alpha$ behavior of the chirality, shown in \cref{finiteT}, which is linear in $\alpha$ and quadratic in $T$.
The $\sim T^2$ behavior can also be obtained by the method of Sommerfeld expansion\cite{ashcroft2022solid} of \cref{chiTIntegral}.

\section{The anisotropic XXZ chain with chirality} 
\label{XXZsect}

In this section, we generalize the obtained results to the anisotropic XXZ model defined by the Hamiltonian  
$H_\Delta(J,\alpha)$ in \cref{XXZ}. Even though the XXZ and the chirality terms in \cref{XXZ} define two separate exactly solvable model possessing an infinite number of conserved currents, their sum does not generally lead to an integrable system. The integrability is brought in when the two terms commute for finite $J$ and $\alpha$. This is achieved only in the isotropic limit, $\Delta \to 1$, admitting the Bethe ansatz solution discussed in the previous section.
While the anisotropic XXZ chain with chiral term \cref{XXZ} is not integrable, it is possible to construct a more complicated anti-symmetric three-body chiral interaction which would ensure the integrability in the anisotropic case\cite{tsvelik1990incommensurate,arnaudon2000integrable}. However, we restrict ourselves to the simple, isotropic chirality interaction and treat the model \cref{XXZ} numerically.

\begin{figure}
    \centering
    \hfill
    \begin{minipage}{0.45\textwidth}
    \includegraphics[width=\textwidth]{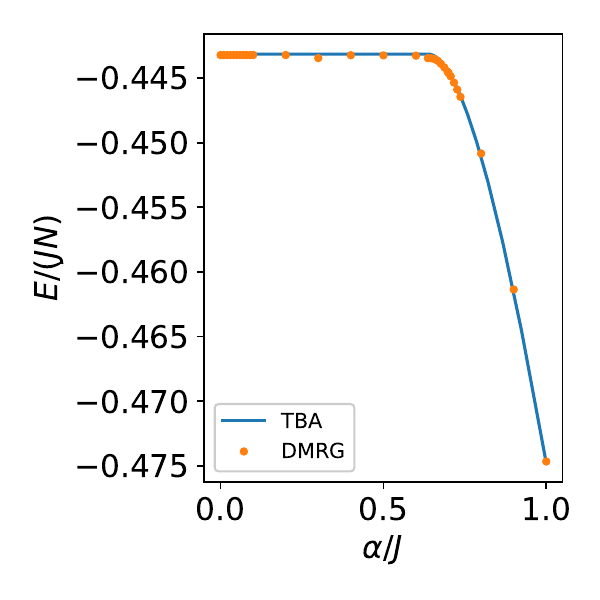}
    \begin{picture}(0,0)
    {\def\unitlength{}
    \ifTwocolumn
        \put(-0.5\linewidth,1.0\linewidth){$(a)$}
    \else
        \put(0\linewidth,1.0\linewidth){$(a)$}
    \fi
    }
    \end{picture}
    \phantomsubcaption{\label{Ealpha}}
    \end{minipage}
    \hfill
    \begin{minipage}{0.45\textwidth}
    \includegraphics[width=\textwidth]{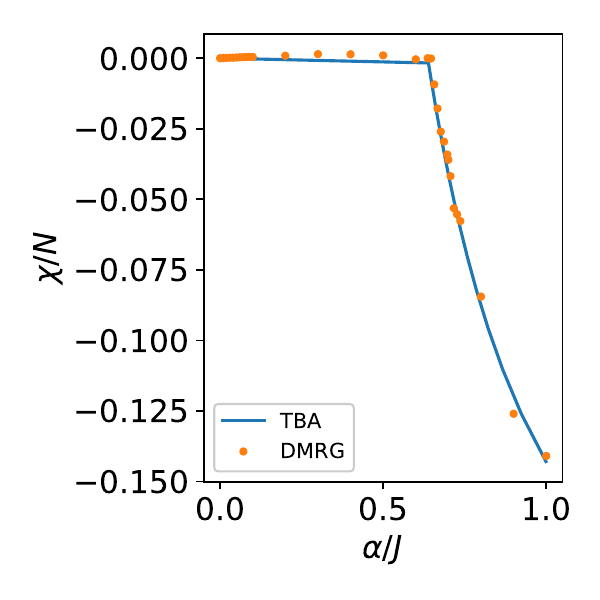}
    \begin{picture}(0,0)
    {\def\unitlength{}
    \ifTwocolumn
        \put(-0.5\linewidth,1.0\linewidth){$(b)$}
    \else
        \put(0\linewidth,1.0\linewidth){$(b)$}
    \fi
    }
    \end{picture}
    \phantomsubcaption{\label{chialpha}}
    \end{minipage}
    \hfill
    \begin{minipage}{0\linewidth}
    \end{minipage}
    \caption{\label{transition} 
The energy (panel \protect\subref{Ealpha}) and the chirality (panel \protect\subref{chialpha}) of the chiral Heisenberg chain. The solid lines are computed from the TBA analytical result, and the dots correspond to the DMRG simulations. One can see the transition (discontinuity) in chirality at $\alpha_c = 2J/\pi$. The energy (chirality) is symmetric (anti-symmetric) around $\alpha=0$ Heisenberg point, so the region with negative $\alpha$ is not shown.}
\end{figure}
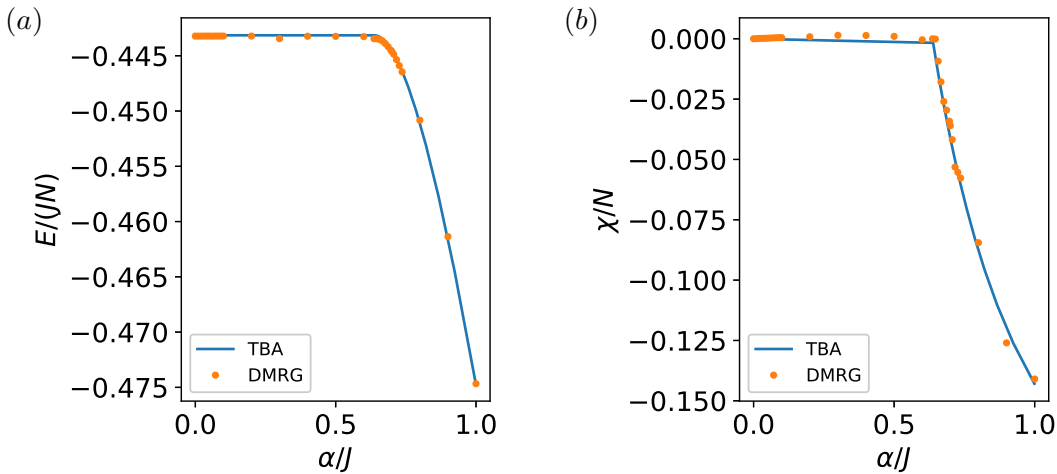

To this end, we employ the DMRG method\cite{tenpy}, allowing us to study portions of the phase diagram \cref{phaseDiagram} with $\Delta\neq 1$.
To classify various phases, we computed a set of order parameters. In particular, the parameter
$M^{+} = \langle s^z_n \rangle$ is computed for the ferromagnetic order,
$M^{-} = \langle (-1)^n s^z_n \rangle$ for the antiferromagnetic order, and 
$\chi= \langle\mathbf{s}_n \cdot (\mathbf{s}_{n+1} \cross \mathbf{s}_{n+2}) \rangle$ -- for the chiral order.
At $\Delta = -1$, we see that even small $\alpha$ already establishes the chiral order. As $\Delta$ decreases further, the critical $\alpha$ between the ferromagnetic and the chiral phases gets larger (see the red curves in \cref{phase_diagrams_chiral}).
On the opposite direction, when $\Delta \gtrsim 1$, with increasing $\alpha$, the chiral Heisenberg chain exhibits antiferromagnetic order for small $\alpha$, paramagnetic order for intermediate $\alpha$, and enters into the chiral phase for larger $\alpha$.

\section{Finite size scaling}\label{finite_size_scaling}

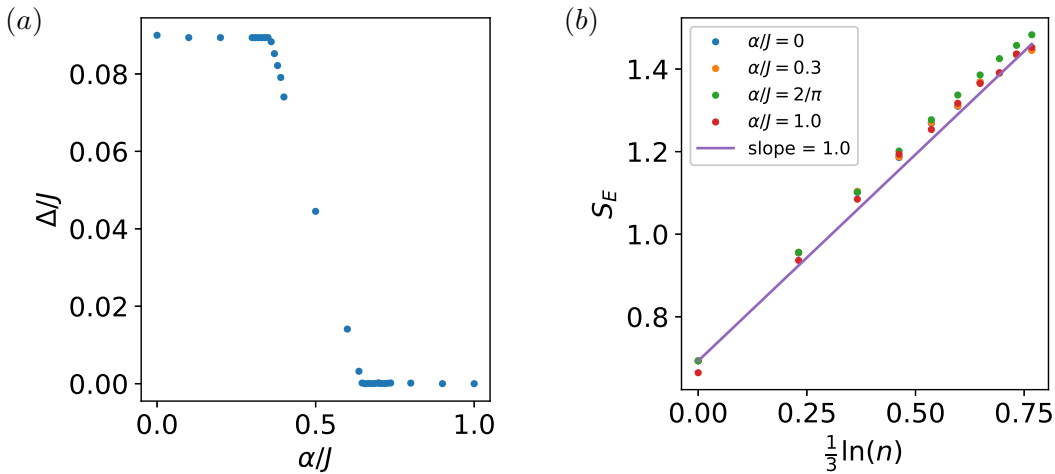
\begin{figure}
    \centering
    \hfill
    \begin{minipage}{0.45\textwidth}
    \includegraphics[width=\textwidth]{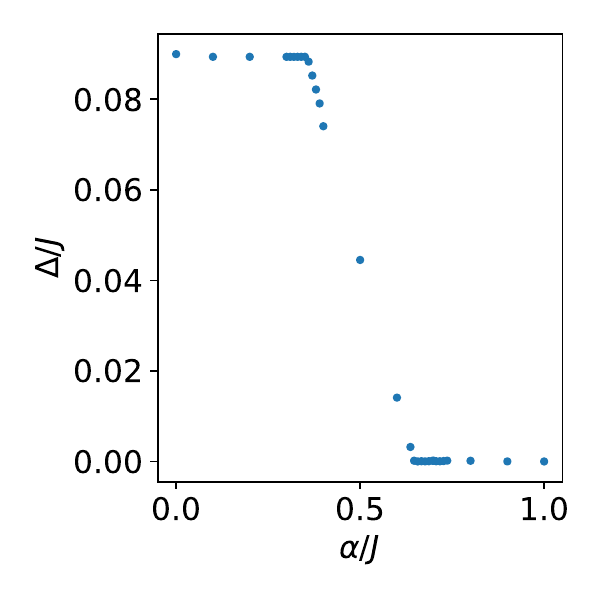}
    \begin{picture}(0,0)
    {\def\unitlength{}
    \ifTwocolumn
        \put(-0.5\linewidth,1.0\linewidth){$(a)$}
    \else
        \put(0\linewidth,1.0\linewidth){$(a)$}
    \fi
    }
    \end{picture}
    \phantomsubcaption{\label{gap}}
    \end{minipage}
    \hfill
    \begin{minipage}{0.45\textwidth}
    \includegraphics[width=\textwidth]{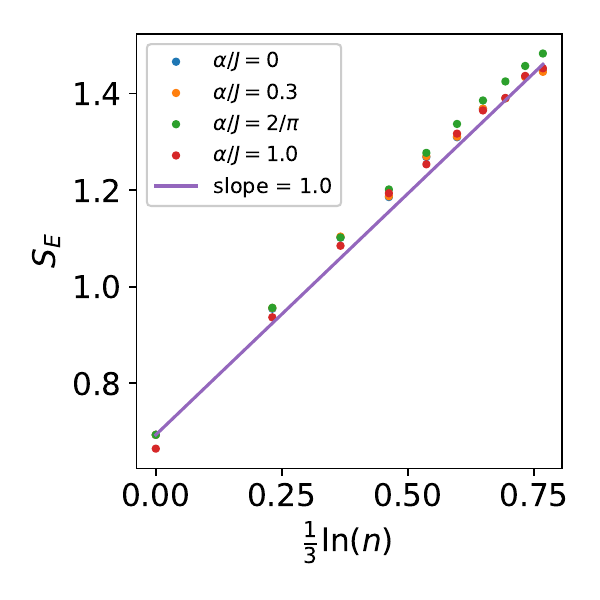}
    \begin{picture}(0,0)
    {\def\unitlength{}
    \ifTwocolumn
        \put(-0.5\linewidth,1.0\linewidth){$(b)$}
    \else
        \put(0\linewidth,1.0\linewidth){$(b)$}
    \fi
    }
    \end{picture}
    \phantomsubcaption{\label{S_E}}
    \end{minipage}
    \hfill
    \begin{minipage}{0\linewidth}
    \end{minipage}
    \caption{\label{gapless}
Panel \protect\subref{gap} shows the excitation gap as a function of $\alpha$. The finite-size effect, giving a finite energy gap in the Heisenberg phase, suppresses upon crossing the transition point $\alpha_c = 2J/\pi$. Panel \protect\subref{S_E} shows the entanglement entropy as a function of the logarithmic subsystem size. The slope gives the central charge of the emergent CFT, $c=1$.
    }
\end{figure}

In this section, the DMRG calculation\cite{tenpy} is performed for the system with periodic boundary conditions to further understand the finite size corrections to various characteristics of the chiral anisotropic Heisenberg (XXZ) model \cref{XXZ}. Both the ground state energy and the chirality per spin agree perfectly with the TBA results, as shown in \cref{transition}. The phase for $\alpha < 2J/\pi$ is in the same universality class as the Heisenberg $c=1$ CFT, whose gap at finite size (with an even number of total spins) decreases as $1/N$ upon increasing the system size, $N$. For $\alpha > 2J/\pi$, the gap closes because the combination of time reversal $\mathcal{T}$ and parity reversal $\mathcal{P}$ leaves the Hamiltonian invariant. Hence, $\mathcal{PT}$ relates the degenerate ground states.

In general, the entanglement entropy of a (1+1)D CFT with periodic boundary conditions is given by the universal formula\cite{calabrese2009entanglement}
\begin{equation}\begin{aligned}
    S_E&=\frac{c}{3}\log\left(\frac{N}{\pi}\sin\frac{\pi n}{N}\right)+c_1^{\prime},
\end{aligned}\end{equation}
where $c$ is the central charge, 
$n$ is the subsystem size, and $c_1'$ is a non-universal constant. In our case of the chiral Heisenberg model, 
both states have the same central charge $c=1$, which can be extracted from the small $n<N$ scaling of the entanglement entropy,  $S_E = \frac{c}{3} \ln(n) + c_1'$. 
The corresponding numerical data is shown in \cref{gapless}.
At maximally entangled $n=N/2$ case, the entanglement entropy as a function of $\alpha$ exhibits a transition: starting at $\frac{1}{3}\ln(N/\pi) + c_1'$ for $\alpha=0$ in agreement with the $c=1$ Heisenberg CFT, it remains constant for up to $\alpha \approx 2J/\pi$. Then it decreases to a smaller value 
before reverting to $\frac{1}{3}\ln(N/\pi)$ for significantly larger values of $\alpha \gg 2J/\pi$.


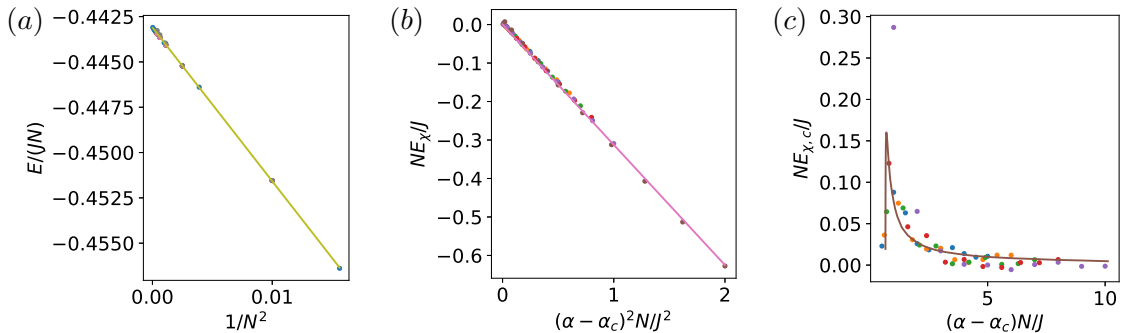
\begin{figure}
    \centering
    \hfill
    \begin{minipage}{0.32\textwidth}
    \includegraphics[width=\textwidth]{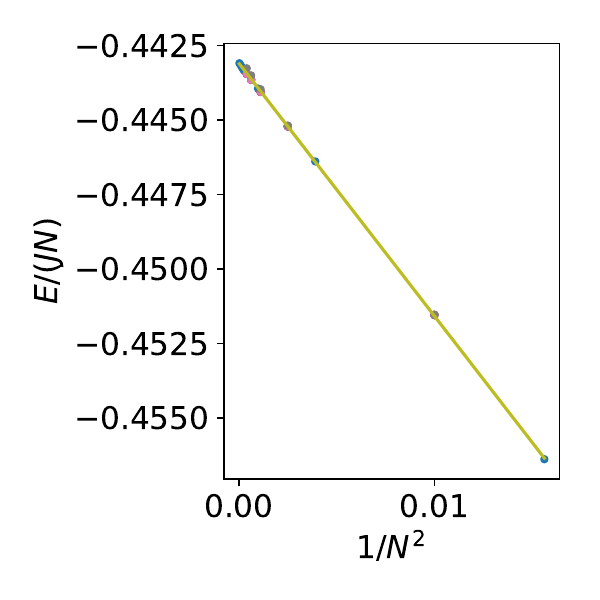}
    \begin{picture}(0,0)
    {\def\unitlength{}
    \ifTwocolumn
        \put(-0.5\linewidth,1.0\linewidth){$(a)$}
    \else
        \put(0\linewidth,1.0\linewidth){$(a)$}
    \fi
    }
    \end{picture}
    \phantomsubcaption{\label{E_over_N}}
    \end{minipage}
    \hfill
    \begin{minipage}{0.32\textwidth}
    \includegraphics[width=\textwidth]{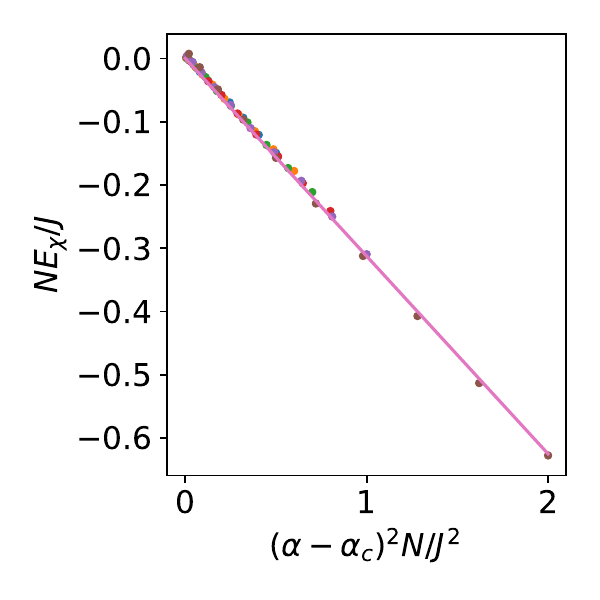}
    \begin{picture}(0,0)
    {\def\unitlength{}
    \ifTwocolumn
        \put(-0.5\linewidth,1.0\linewidth){$(b)$}
    \else
        \put(0\linewidth,1.0\linewidth){$(b)$}
    \fi
    }
    \end{picture}
    \phantomsubcaption{\label{E_chi}}
    \end{minipage}
    \hfill
    \begin{minipage}{0.32\textwidth}
    \includegraphics[width=\textwidth]{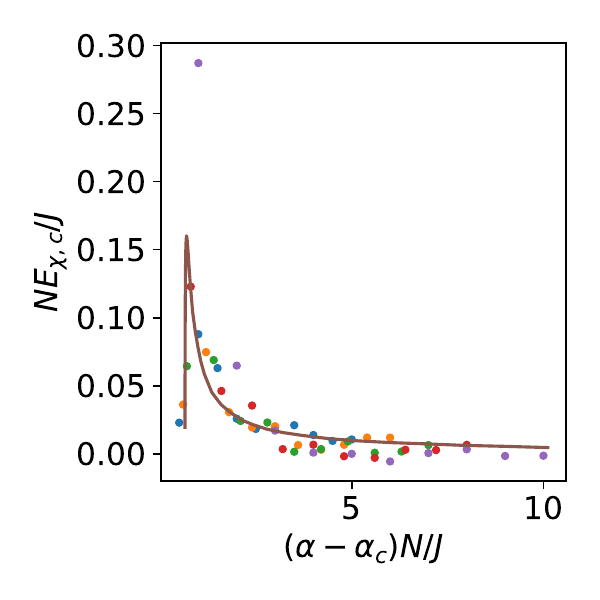}
    \begin{picture}(0,0)
    {\def\unitlength{}
    \ifTwocolumn
        \put(-0.5\linewidth,1.0\linewidth){$(c)$}
    \else
        \put(0\linewidth,1.0\linewidth){$(c)$}
    \fi
    }
    \end{picture}
    \phantomsubcaption{\label{E_chi_c}}
    \end{minipage}
    \hfill
    %
    \caption{\label{correction}
Panel \protect\subref{E_over_N} shows the $E/N$ as a function of $1/N^2$, the linear behavior indicates the finite size scaling \cref{EfiniteHeisenberg}.
Panel \protect\subref{E_chi} and \protect\subref{E_chi_c} show the energy of only the chiral term and its finite size correction. The chiral energy shows a quadratic behavior.
    }
\end{figure}

For $\alpha < 2J/\pi$, the chiral order parameter vanishes, and the model is effectively in the same universality class as the isotropic Heisenberg Hamiltonian. Hence, expecting the energy to have the same scaling behavior as the Heisenberg CFT is natural. So, we indeed have that
\begin{equation}\label{EfiniteHeisenberg}
    E / J = E_0 N - \frac{\pi^2}{12N}+O\left(\frac{1}{N\ln N}\right),
\end{equation}
as shown in \cref{E_over_N}, where $E_0 = 1/4 - \ln 2$.

For $\alpha > 2J/\pi$, around the transition point, we have 
\begin{equation}
    E / J = \epsilon N + \frac{cv}{N} \left[g(0,\zeta (\alpha-\alpha_c) N) - \frac{\pi}{6}\right]+O\left(\frac{1}{N\ln N}\right), 
\end{equation}
where the $\epsilon = 1/4 - \ln 2 - \kappa (\alpha/J - 2/\pi)^2$. The constant $\kappa \approx 0.370154$ can be derived from the TBA solution, see \cref{closedform}. The DMRG result also agrees with this, as shown in \cref{E_chi}. Within the numerical precision, the shape of the scaling function $g(0,\zeta (\alpha-\alpha_c) N)$ agrees with the universal scaling function in time-reversal symmetry broken criticalities of \cref{Egni}, see \cite{wang2020universal}. More precisely, it agrees with the universal behavior  observed in non-interacting systems \cite{wang2020universal}, when the gap, $m \to 0$, while the expression of the symmetry-breaking scaling variable is different because of the strong interaction.

\section{Conclusion and outlook} \label{conclusion}

One of the motivations of the present study comes from the desire to understand the formation mechanism of chiral spin liquids, long-range entangled quantum states of matter lacking the magnetic order while supporting the topological order with spontaneous breaking of the time-reversal symmetry\cite{kalmeyer1987equivalence,kalmeyer1989theory,wen1989chiral}. Interestingly, the strong scalar spin-$1/2$ chiral interaction between the spins on sites on triangles of the kagome lattice, $\sum_\Delta \mathbf{s}_i \cdot (\mathbf{s}_j \cross \mathbf{s}_k)$, can drive the two-dimensional spin system to a chiral spin liquid state\cite{bauer2014chiral,kumar2015chiral,niu2022chiral,claassen2017dynamical,schweika2022chiral}. The question still remains: what is the mechanism of the formation of the chiral spin liquid state in an array of individual XXZ models upon the adiabatic increase of the interchain couplings, and how can one enter the chiral spin liquid phase? A related open question is how can the chiral spin liquid state be formed spontaneously in 2D without explicitly breaking the time-reversal symmetry. 

To address these questions, we, as the first step,  studied the chiralization of the Heisenberg chain with an additional pseudo-scalar chiral interaction, a conserved current of the Heisenberg model. We showed that the ground state of this system maintains a zero chirality until it reaches a transition point at $\alpha=\alpha_c = 2J / \pi$. Both of these phases are gapless and can be described by CFTs with an identical central charge of one. Their ground state or lightest weight primary fields transform differently under $\mathcal{PT}$ symmetry, making them symmetry-enriched CFTs. At low temperatures, the chirality develops as long as the chiral term is turned on, and the chirality exhibits a quadratic temperature dependence. This intriguing transition can be explored by considering the finite-size effects around the transition point. The linear in $N$ term in the ground state energy shows  $\propto\alpha^2$ dependence. The $\sim 1/N$ ground state energy correction above the transition point in this system follows a universal function exhibiting non-monotonic behavior that detects the transition playing the role of the order parameter.

As reported in the present work, it would be interesting to experimentally observe the chiral phase and the transition into it. The experimental study of the Heisenberg chain in electronic systems has been extensively explored \cite{steiner1976theoretical,mourigal2013fractional}. More recently, in the optical lattice Hubbard model, the Heisenberg chain with tunable parameters has attracted significant attention \cite{jepsen2020spin,jepsen2021transverse}. Notably, the pseudo-scalar chiral interaction can emerge from the $t/U$ expansion of the half-filled Hubbard model, a phenomenon studied in depth in various works \cite{takahashi1977half,macDonald1988expansion,motrunich2006orbital}.
These possible experiment setups facilitate the experimental investigation of the chiral Heisenberg chain presented in this theoretical paper. This accessibility promises to unlock exploration into both chiral spin liquids and symmetry-enriched quantum criticality.

\begin{figure}
    \centering
    \begin{minipage}{0.7\textwidth}
    \includegraphics[width=\textwidth]{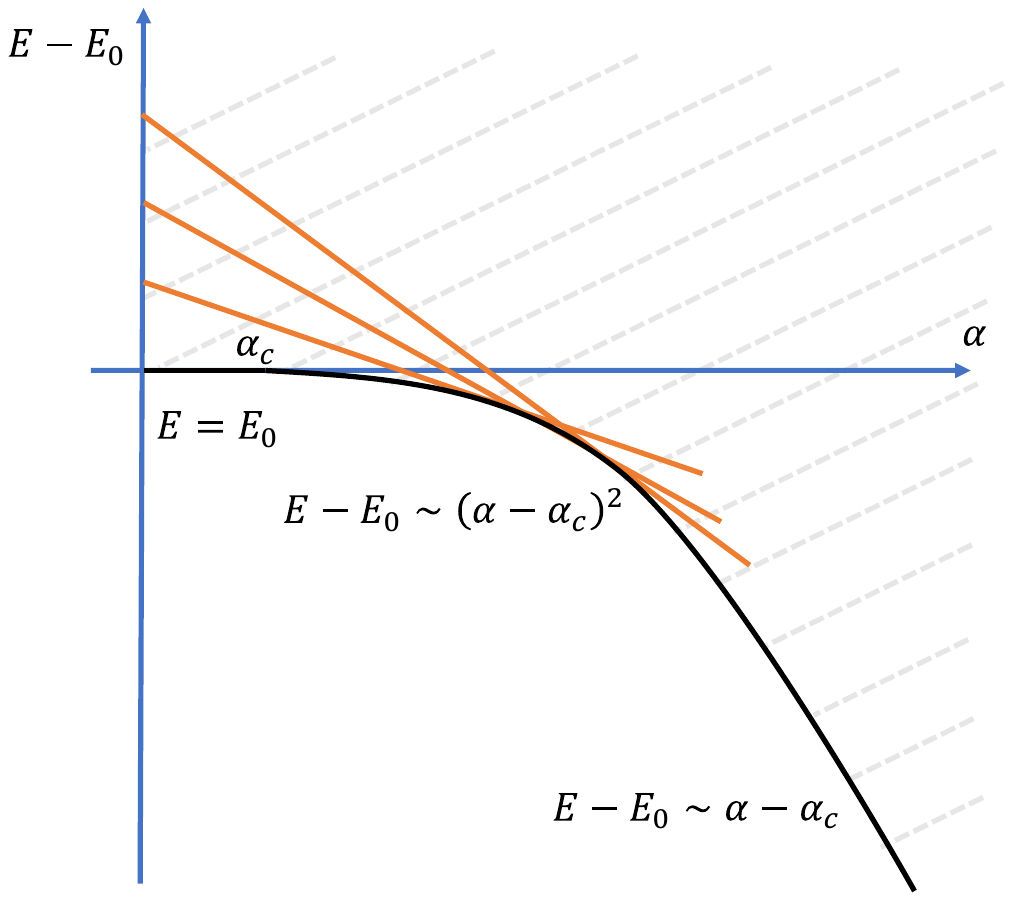}
    \end{minipage}
    \caption{The schematic of the spectrum of the chiral Heisenberg chain. The energy is shifted by the ground state energy of isotropic Heisenberg chain $E_0=-J(\ln 2-1/4)$.
    The solid black line represents the ground state energy, while the dashed gray lines show the excitation spectrum.
    Starting from a state of the Heisenberg chain, the chiral term will decrease its energy linearly in $\alpha$, turning an excited state at zero $\alpha$ into the ground state at higher $\alpha$, shown by orange lines.
    The quadratic energy behavior in the vicinity of $\alpha_c$ is a result of many crossing points of Heisenberg excited states.}
    \label{fig:schematic}
\end{figure}

Although the two critical phases can be described by symmetry-enriched CFTs, it would be interesting to study the properties of the critical point between these two critical phases further. One of the possibilities is the Lifshitz point, which is indicated by its multi-criticality nature and the quadratic ground state energy scaling $E \sim (\alpha-\alpha_c)^2$ (see \cref{fig:schematic} for the schematic energy spectrum) in the vicinity of $\alpha_c$. Another interesting property is the relation to the commensurate - incommensurate criticality discussed by Tsvelik\cite{tsvelik1990incommensurate}. The possible connection to the Lifshitz point at multi-criticality is another open question\cite{howes1983quantum,howes1983commensurate,chepiga2021lifshitz}.
It would also be useful to have quantum field theory description of such critical point with asymmetric (chiral) dispersion.


Another interesting direction for future research is the study of the spin dynamics in the chiral XXZ chain, which remains an open problem. Investigating the spin dynamics within the framework of the present model involves a comprehensive analysis of the impact of the chiral term on the hydrodynamic approach and the resulting alteration in the universality class (dynamic exponent) of the model. Furthermore, the distinct symmetry-enriched CFTs describing the critical Heisenberg and the chiral phases differ in their lightest primary fields, suggesting an anticipated modification in the corresponding Knizhnik-Zamolodchikov equations for the correlation functions of primary fields\cite{sedrakyan2022quantum}.

\begin{acknowledgments}
We thank Fabian Essler and Alex Kamenev for their insightful comments. C. W. and T. S. gratefully acknowledge support from the Simons Center for Geometry and Physics, Stony Brook University, at which some of the research for this paper was performed.
\end{acknowledgments}

\appendix

\section{Transfer matrix method for the chiral Heisenberg chain}
\label{dlnT}

In this appendix, we will review the transfer matrix method for the Heisenberg chain and derive the chirality operator as the conserved quantity of the Heisenberg model from the logarithmic derivative of the transfer matrix,
\begin{dmath} \label{chi-dlnT}
     \mathcal{C} = \left. -\frac{i}{4} \frac{d^2}{du^2} \ln T(u) \right|_{u=i/2}
     +\frac{1}{4} iL.
\end{dmath}
This will reproduce the discussion presented in the introduction of the main text (see below \cref{XXZ}).

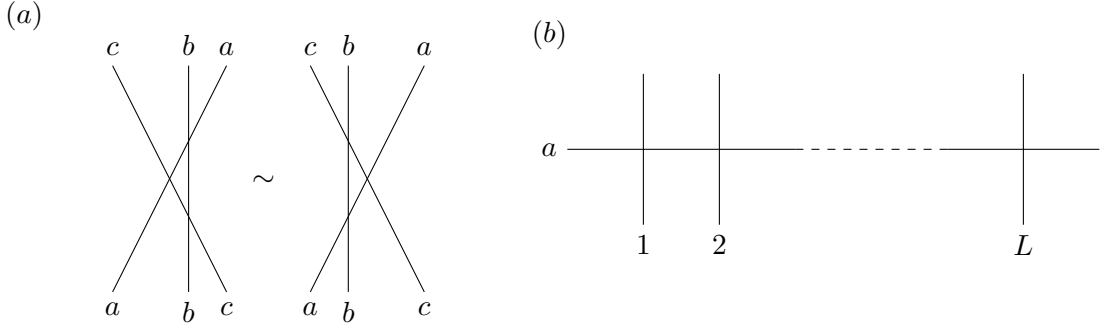
\begin{figure}[h]
    \centering
    \begin{minipage}{0.45\textwidth}
    \begin{picture}(0,0)
    {\def\unitlength{}
    \put(0\textwidth,0\textwidth){$(a)$}}
    \end{picture}
    \begin{equation*}
        \vcenter{\hbox{
        \begin{tikzpicture}
            \draw (0,0) node[anchor=north] {$a$} -- (1.5,3) node[anchor=south] {$a$};
            \draw (1.5,0) node[anchor=north] {$c$} -- (0,3) node[anchor=south] {$c$};
            \draw (1,0) node[anchor=north] {$b$} -- (1,3) node[anchor=south] {$b$};
        \end{tikzpicture}}}
        \sim
        \vcenter{\hbox{
        \begin{tikzpicture}
            \draw (0,0) node[anchor=north] {$a$} -- (1.5,3) node[anchor=south] {$a$};
            \draw (1.5,0) node[anchor=north] {$c$} -- (0,3) node[anchor=south] {$c$};
            \draw (0.5,0) node[anchor=north] {$b$} -- (0.5,3) node[anchor=south] {$b$};
        \end{tikzpicture}}}
    \end{equation*}
    \phantomsubcaption{\label{Rmatrix}}
    \end{minipage}
    \begin{minipage}{0.45\textwidth}
    \begin{picture}(0,0)
    {\def\unitlength{}
    \put(0\textwidth,0\textwidth){$(b)$}}
    \end{picture}
    \begin{equation*}
        \begin{tikzpicture}
            \draw (0,0) node[anchor=east] {$a$} -- (3,0);
            \draw (1,-1) node[anchor=north] {$1$} -- (1,1);
            \draw (2,-1) node[anchor=north] {$2$} -- (2,1);
            \draw[dashed] (3,0) -- (5,0);
            \draw (5,0) -- (7,0);
            \draw (6,-1) node[anchor=north] {$L$} -- (6,1);
        \end{tikzpicture}
    \end{equation*}
    \phantomsubcaption{\label{Monodromy}}
    \end{minipage}
    \caption{Graphic representation of the Yang-Baxter equation \protect\subref{Rmatrix} and the monodromy matrix \protect\subref{Monodromy}. Each cross represents an $R$-matrix.
    The indices $a$, $b$ and $c$ live in the $SU(2)$ auxiliary space, while indices $1,2,\dots$ are site indices.}
    \label{fig:YBEM}
\end{figure}

Below we will list the standard steps of the construction of the transfer matrix for the Heisenberg spin chain with the Hamiltonian
\begin{dmath}
    H =
    J \sum_{n} \mathbf{s}_n \cdot \mathbf{s}_{n+1}.
\end{dmath}
Here, the spin-1/2 operators are half of the Pauli matrices, $s^\mu=\frac{\hbar}{2}\sigma_\mu, \mu=x,y,z$,
can be rewritten in terms of the permutation operator:
\begin{dmath} \label{Pab}
    P_{ab}=\frac12(I + \sum_{i=x,y,z} \sigma_{a}^{i}\otimes\sigma_{b}^{i}).
\end{dmath}
The Heisenberg Hamiltonian then acquires the following form:
\begin{dmath}
    H =
    \frac{J}{2} \sum_{n} P_{n,n+1} - \frac{JL}{4}.
\end{dmath}
As the next step, we identify the $R$-matric of the model by first  defining the Lax operator as
\begin{dmath} \label{lax}
    L_{na}(u)=uI + \frac{i}{2} \sum_{i=x,y,z} \sigma_{n}^{i}\otimes\sigma_{a}^{i},
\end{dmath}
where the first index $n$ acts on spin at site $n$ and the second index acts on an auxiliary $SU(2)$ space $V$.
We are here using the notations from Ref.~\onlinecite{faddeev1996algebraic}.
The $R_{ab}$ matrix, which acts on $V \otimes V$, is 
\begin{dmath}
    R_{ab} = u I + i P_{ab}.
\end{dmath}
It is important to emphasize here that this $R_{ab}$ matrix satisfies the Yang-Baxter equation
\begin{dmath}
    R_{ab}(u_1-u_2)R_{ac}(u_1-u_3)R_{bc}(u_2-u_3)
    = R_{bc}(u_2-u_3)R_{ac}(u_1-u_3)R_{ab}(u_1-u_2),
\end{dmath}
as illustrated in \cref{Rmatrix}. This equation is the defining property of the (Yang-Baxter) integrability.
The related objets are the transfer matrix $T$ and the monodromy matrix $M_a$.
The transfer matrix can be defined as
\begin{dmath}
    T(u) = \tr_a M_{a}(u)
\end{dmath}
where $M_{a}(u)$ is the monodromy matrix given by the following analytical expression (shown graphically in \cref{Monodromy}):
\begin{dmath}
    M_{a}(u) = L_{1a}(u)L_{2a}(u)\dots L_{La}(u).
\end{dmath}

Having the definitions above, one can generate an infinite number of conserved quantities (in the thermodynamic limit $N\rightarrow \infty$) upon taking the derivatives of the logarithm of the transfer matrix.
We will compute up to the second-order derivative of the transfer matrix in the following
to get total energy and chirality.

To this end, we notice that the Lax operator \cref{lax}
is linear in $u$, so all the higher order derivatives, beyond the first, of the Lax operator vanish.
So the first and the second-order derivative of the transfer matrix are separately evaluated to be
\begin{dmath}
    \frac{d T(u)}{du} = \sum_{n} \tr_a( L_{1a}(u)L_{2a}(u) \dots \check{L}_{na}(u) \dots L_{La}(u) )
\end{dmath}
and
\begin{dmath}
    \frac{d^2 T(u)}{du^2} = \sum_{m \neq n} \tr_a( L_{1a}(u)L_{2a}(u) \dots \check{L}_{ma}(u) \dots \check{L}_{na}(u) \dots L_{La}(u) ).
\end{dmath}
Here the ``check" mark on the operator, $\check{L}$, indicates that operator $L$ is not in the product.
At $u=i/2$, the Lax operator becomes the permutation operator, $L_{mn}(i/2) = iP_{mn}$.
It is straightforward to verify that $P_{mn}$ exchanges the spin states on the sites $m$ and $n$ and satisfies $P_{ma}P_{na} = P_{na}P_{mn}$.
So, the transfer matrix at $u=i/2$ becomes
\begin{dmath}
    T(i/2) = i^L P_{L,L-1} P_{L-1,L-2} \dots P_{2,1},
\end{dmath}
and its inverse is
\begin{dmath} \label{T-1}
    T^{-1}(i/2) = (-i)^L P_{1,2} P_{2,3} \dots P_{L-1,L}.
\end{dmath}

The derivatives of the transfer matrix become
\begin{dmath} \label{dTdu}
    \frac{d T(i/2)}{du} = i^{L-1} \sum_{n} P_{L,L-1} P_{L-1,L-2} \dots P_{n+1,n-1} \dots P_{2,1},
\end{dmath}
and
\begin{dmath} \label{d2Tdu2}
    \frac{d^2 T(i/2)}{du^2} = i^{L-2} \sum_{m \neq n} P_{L,L-1} P_{L-1,L-2} \dots P_{m+1,m-1} \dots P_{n+1,n-1} \dots P_{2,1}.
\end{dmath}
Combining \cref{T-1,dTdu,d2Tdu2}, one finds
\begin{dmath}
    \frac{d \ln T(i/2)}{du} = -i \sum_n P_{n,n+1},
\end{dmath}
and
\begin{dmath} \label{d2lnTdu2}
    \frac{d^2 \ln T(i/2)}{du^2} = \sum_n P_{n,n+1}^2 + \sum_n \left( P_{n+1,n}P_{n,n-1}-P_{n-1,n}P_{n,n+1} \right).
\end{dmath}
Now, given the  definition \cref{Pab}, one can verify that
\begin{dmath} \label{PPchi}
    P_{n+1,n}P_{n,n-1}-P_{n-1,n}P_{n,n+1} = 4i \mathcal{C}
\end{dmath}
where $\mathcal{C}$ is the total chirality, \textit{i.e.}, $\mathcal{C}= \sum_n \mathbf{s}_{n} \cdot (\mathbf{s}_{n+1} \cross \mathbf{s}_{n+2})$. With \cref{d2lnTdu2,PPchi}, one arrives at \cref{chi-dlnT} in the main text.

\section{Exact diagonalization of the chiral spin chain}
\label{EDchiral}
In this section, we show the exact diagonalization results for the purely chiral chain with a small total number of sites. The Hamiltonian has the form
\begin{equation}
\begin{aligned}
    H_\Delta(J=0, \alpha=1) =
    \sum_{n=1}^{N} \mathbf{s}_n \cdot (\mathbf{s}_{n+1} \cross \mathbf{s}_{n+2}).
\end{aligned}
\end{equation}
For this model, we present a calculation supporting the conjecture on the solution pattern for general system size, as discussed in the introduction (see around \cref{EDchiral}).

For $N=3$, the ground state is doubly degenerate
\begin{equation}
\begin{aligned}
    \ket{g_1} &=
    \frac{(-1 - i \sqrt{3})}{2} \ket{\uparrow,\downarrow,\downarrow}
    +\frac{(-1 + i \sqrt{3})}{2} \ket{\downarrow,\uparrow,\downarrow}
    + \ket{\downarrow,\downarrow,\uparrow}, \\
    \ket{g_2} &=
    \frac{(-1 + i \sqrt{3})}{2} \ket{\uparrow,\uparrow,\downarrow}
    +\frac{(-1 - i \sqrt{3})}{2} \ket{\uparrow,\downarrow,\uparrow}
    + \ket{\downarrow,\uparrow,\uparrow},
\end{aligned}
\end{equation}
with energy $E_0 = -\frac{3\sqrt{3}}{4}$, and the first excited state is fourfold degenerate
\begin{equation}
\begin{aligned}
    \ket{e_1} &=
    \ket{\downarrow,\downarrow,\downarrow}, \\
    \ket{e_2} &=
    \ket{\uparrow,\downarrow,\downarrow}
    + \ket{\downarrow,\uparrow,\downarrow}
    + \ket{\downarrow,\downarrow,\uparrow}, \\
    \ket{e_3} &=
    \ket{\uparrow,\uparrow,\downarrow} 
    + \ket{\uparrow,\downarrow,\uparrow}
    + \ket{\downarrow,\uparrow,\uparrow}, \\
    \ket{e_4} &= 
    \ket{\uparrow,\uparrow,\uparrow}
\end{aligned}
\end{equation}
with energy $E_1 = 0$.

For $N=4$, the ground state is fourfold degenerate
\ifTwocolumn
    \begin{equation}
    \begin{aligned}
        \ket{g_1} &=
        -i \ket{\uparrow,\downarrow,\downarrow,\downarrow}
        - \ket{\downarrow,\uparrow,\downarrow,\downarrow}
        +i \ket{\downarrow,\downarrow,\uparrow,\downarrow}
        + \ket{\downarrow,\downarrow,\downarrow,\uparrow}, \\
        \ket{g_2} &=
        - \ket{\uparrow,\uparrow,\downarrow,\downarrow}
        -i \ket{\uparrow,\downarrow,\downarrow,\uparrow}
        + \ket{\downarrow,\downarrow,\uparrow,\uparrow}
        +i \ket{\downarrow,\uparrow,\uparrow,\downarrow}, \\
        \ket{g_3} &=
        i \ket{\uparrow,\uparrow,\uparrow,\downarrow}
        - \ket{\uparrow,\uparrow,\downarrow,\uparrow}
        -i \ket{\uparrow,\downarrow,\uparrow,\uparrow}
        + \ket{\downarrow,\uparrow,\uparrow,\uparrow}, \\
        \ket{g_4} &=
        - \ket{\uparrow,\uparrow,\downarrow,\downarrow}
        + \ket{\uparrow,\downarrow,\uparrow,\downarrow}
        + \ket{\downarrow,\uparrow,\downarrow,\uparrow} \\
        &+ \frac{-1-i}{2} \ket{\uparrow,\downarrow,\downarrow,\uparrow}
        + \frac{-1+i}{2} \ket{\downarrow,\uparrow,\uparrow,\downarrow}
    \end{aligned}
    \end{equation}
\else
    \begin{equation}
    \begin{aligned}
        \ket{g_1} &=
        -i \ket{\uparrow,\downarrow,\downarrow,\downarrow}
        - \ket{\downarrow,\uparrow,\downarrow,\downarrow}
        +i \ket{\downarrow,\downarrow,\uparrow,\downarrow}
        + \ket{\downarrow,\downarrow,\downarrow,\uparrow}, \\
        \ket{g_2} &=
        - \ket{\uparrow,\uparrow,\downarrow,\downarrow}
        -i \ket{\uparrow,\downarrow,\downarrow,\uparrow}
        + \ket{\downarrow,\downarrow,\uparrow,\uparrow}
        +i \ket{\downarrow,\uparrow,\uparrow,\downarrow}, \\
        \ket{g_3} &=
        i \ket{\uparrow,\uparrow,\uparrow,\downarrow}
        - \ket{\uparrow,\uparrow,\downarrow,\uparrow}
        -i \ket{\uparrow,\downarrow,\uparrow,\uparrow}
        + \ket{\downarrow,\uparrow,\uparrow,\uparrow}, \\
        \ket{g_4} &=
        - \ket{\uparrow,\uparrow,\downarrow,\downarrow}
        + \ket{\uparrow,\downarrow,\uparrow,\downarrow}
        + \ket{\downarrow,\uparrow,\downarrow,\uparrow}
        + \frac{-1-i}{2} \ket{\uparrow,\downarrow,\downarrow,\uparrow}
        + \frac{-1+i}{2} \ket{\downarrow,\uparrow,\uparrow,\downarrow}
    \end{aligned}
    \end{equation}
\fi
with energy $E_0 = -1$, and the first excited state is a threefold degenerate
\begin{equation}
\begin{aligned}
    \ket{e_1} &=
    - \ket{\uparrow,\downarrow,\downarrow,\downarrow}
    + \ket{\downarrow,\uparrow,\downarrow,\downarrow}
    - \ket{\downarrow,\downarrow,\uparrow,\downarrow}
    + \ket{\downarrow,\downarrow,\downarrow,\uparrow}, \\
    \ket{e_2} &=
    - \ket{\uparrow,\downarrow,\uparrow,\downarrow}
    + \ket{\downarrow,\uparrow,\downarrow,\uparrow}, \\
    \ket{e_3} &=
    - \ket{\uparrow,\uparrow,\uparrow,\downarrow}
    + \ket{\uparrow,\uparrow,\downarrow,\uparrow}
    - \ket{\uparrow,\downarrow,\uparrow,\uparrow}
    + \ket{\downarrow,\uparrow,\uparrow,\uparrow},
\end{aligned}
\end{equation}
with energy $E_1 = -\frac12$.
Based on these observations (and checking these for slightly larger values of $N$), we conjecture that the ground state and first excited state are
\begin{equation}
\begin{aligned}
    \ket{g_1} &=
    \sum_n \exp(\frac{2\pi i n}{N}) \ket{\downarrow,\downarrow, ..., \uparrow_{{\rm site~} n}, ..., \downarrow,\downarrow}, \\
    \ket{g_2} &=
    \sum_n \exp(-\frac{2\pi i n}{N}) \ket{\uparrow,\uparrow, ..., \downarrow_{{\rm site~} n}, ..., \uparrow,\uparrow}, \\
    \ket{e_1} &=
    \sum_n (-1)^{\Pi[n(N+1)]} \ket{\downarrow,\downarrow, ..., \uparrow_{{\rm site~} n}, ..., \downarrow,\downarrow}, \\
    \ket{e_2} &=
    \sum_n (-1)^{\Pi[n(N+1)]} \ket{\uparrow,\uparrow, ..., \downarrow_{{\rm site~} n}, ..., \uparrow,\uparrow},
\end{aligned}
\end{equation}
where $\Pi: \mathbb{Z}\rightarrow\{0,1\}$ is the parity of the argument.
$\ket{g_1}$ and $\ket{g_2}$ are related by the combination of parity and time reversal $\ket{g_1} = \mathcal{PT} \ket{g_2}$, so are $\ket{e_1}$ and $\ket{e_2}$, \textit{i.e.}, $\ket{e_1} = \mathcal{PT} \ket{e_2}$.
 We suggest that the ground state is a superposition of a single up spin in a down spin background (down spin in an up spin background) at all possible sites with a twisting phase.
We numerically checked this conjecture for a system of size $N=6$.

The primary field is related to the operator connecting the ground state and the low-energy excitation,
\begin{equation} 
p \equiv \prod_{n=1}^L P_n\left(\frac{2\pi n}{N}-\pi \Pi\left[n(N + 1)\right]\right),
\end{equation}
where $P_n(\varphi)$ is the phase shift gate on site $n$ that maps the basis states $\ket{\downarrow} \to \ket{\downarrow}$ and $\ket{\uparrow} \to e^{i\varphi}\ket{\uparrow}$. 
This operator is introduced in \cref{phasetwist} of the main text.

\section{Bosonization of the chiral Heisenberg chain}\label{BRG}

In this section, we discuss the bosonization of the chiral Heisenberg chain.
Following the standard steps of the bosonization scheme, one can perform the Jordan-Wigner transformation, take the continuous limit, and then bosonize the Heisenberg chain:
\ifTwocolumn
    \begin{equation}
    \begin{aligned}
        H_0 &= J/2 (s_n^+ s_{n+1}^- + h.c.) + J s_n^{z} s_{n+1}^{z} \\
        &= J/2 (c_n^{\dagger} c_{n+1} + h.c.) + J (n_n-\frac12) (n_{n+1}-\frac12) \\
        &\sim i J \int dx (\psi_R^{\dagger} \partial_x \psi_R - \psi_L^{\dagger} \partial_x \psi_L) \\
        &\qquad\qquad + J \int dx (\rho_R^2 +\rho_L^2 + 4\rho_L \rho_R) \\
        &\sim \frac{v}{2} \int dx [ g^{-1} (\partial_x \phi)^2 + g (\partial_x \theta)^2],
    \end{aligned}
    \end{equation}
\else
    \begin{equation}
    \begin{aligned}
        H_0 &= J/2 (s_n^+ s_{n+1}^- + h.c.) + J s_n^{z} s_{n+1}^{z} \\
        &= J/2 (c_n^{\dagger} c_{n+1} + h.c.) + J \left(n_n-\frac12\right) \left(n_{n+1}-\frac12\right) \\
        &\sim i J \int dx (\psi_R^{\dagger} \partial_x \psi_R - \psi_L^{\dagger} \partial_x \psi_L) 
        + J \int dx (\rho_R^2 +\rho_L^2 + 4\rho_L \rho_R) \\
        &\sim \frac{v}{2} \int dx [ g^{-1} (\partial_x \phi)^2 + g (\partial_x \theta)^2],
    \end{aligned}
    \end{equation}
\fi
where
\begin{equation}\begin{aligned}
    \psi_{R}&=\frac{1}{\sqrt{2\pi}}e^{i\sqrt{4\pi} \varphi_{R}} \\
    \psi_{L}&=\frac{1}{\sqrt{2\pi}}e^{-i\sqrt{4\pi} \varphi_L},
\end{aligned}\end{equation}
and $\rho_{R/L} = :\psi_{R/L}^{\dagger} \psi_{R/L}:$. Here the umklapp terms are ignored.
We have also introduced
\begin{equation}\begin{aligned}
    \phi = \varphi_{R} + \varphi_{L} \\
    \theta= \varphi_{R} -\varphi_{L}
\end{aligned}\end{equation}
to write the Hamiltonian in the boson and dual boson representation.

Following the same steps, one can bosonize the chiral term
\ifTwocolumn
    \begin{equation}
    \begin{aligned}
        H_{\chi} &= \frac{i \alpha}{2} \sum_n [s_n^z(s_{n+1}^+ s_{n+2}^- - s_{n+1}^- s_{n+2}^+) \\
        &+s_{n+1}^z(s_{n+2}^+ s_{n}^- - s_{n+2}^- s_{n}^+)
        +s_{n+2}^z(s_{n}^+ s_{n+1}^- - s_{n}^- s_{n+1}^+)] \\
        &= \frac{i \alpha}{2} \sum_i [(n_i-\frac12)(c_{i+1}^{\dagger} c_{i+2} - c_{i+2}^{\dagger} c_{i+1}) \\
        &-\frac12 (c_{i+2}^{\dagger} c_{i} - c_{i}^{\dagger} c_{i+2})
        +(n_{i+2}-\frac12)(c_{i}^{\dagger} c_{i+1} - c_{i+1}^{\dagger} c_{i})] \\
        &\sim -\frac{1}{4} i \alpha \int dx  \psi_L^{\dagger}(x) \psi_L(x+2) + \psi_R^{\dagger}(x) \psi_R(x+2) + h.c. \\
        &= -\frac{1}{2} i \alpha \int dx  \psi_L^{\dagger}(x) \partial_x \psi_L(x) + \psi_R^{\dagger}(x) \partial_x \psi_R(x) \\
        &= \frac{1}{2} \alpha \int dx (\partial_x \varphi_R)^2 - (\partial_x \varphi_L)^2 \\
        &= \frac{1}{2} \alpha \int dx \partial_x \phi \partial_x \theta.
    \end{aligned}
    \end{equation}
\else
    \begin{equation}
    \begin{aligned}
        H_{\chi} &= \frac{i \alpha}{2} \sum_n [s_n^z(s_{n+1}^+ s_{n+2}^- - s_{n+1}^- s_{n+2}^+)
        +s_{n+1}^z(s_{n+2}^+ s_{n}^- - s_{n+2}^- s_{n}^+)
        +s_{n+2}^z(s_{n}^+ s_{n+1}^- - s_{n}^- s_{n+1}^+)] \\
        &= \frac{i \alpha}{2} \sum_i \left[\left(n_i-\frac12\right)(c_{i+1}^{\dagger} c_{i+2} - c_{i+2}^{\dagger} c_{i+1})
        -\frac12 (c_{i+2}^{\dagger} c_{i} - c_{i}^{\dagger} c_{i+2})
        +\left(n_{i+2}-\frac12\right)(c_{i}^{\dagger} c_{i+1} - c_{i+1}^{\dagger} c_{i})\right] \\
        &\sim -\frac{1}{4} i \alpha \int dx  \psi_L^{\dagger}(x) \psi_L(x+2) + \psi_R^{\dagger}(x) \psi_R(x+2) + h.c. \\
        &= -\frac{1}{2} i \alpha \int dx  \psi_L^{\dagger}(x) \partial_x \psi_L(x) + \psi_R^{\dagger}(x) \partial_x \psi_R(x) \\
        &= \frac{1}{2} \alpha \int dx (\partial_x \varphi_R)^2 - (\partial_x \varphi_L)^2 \\
        &= \frac{1}{2} \alpha \int dx \partial_x \phi \partial_x \theta.
    \end{aligned}
    \end{equation}
\fi
After Bosonization, $H_{\chi}$ becomes the summation of the product of four $\psi$ fields with up to second-order derivatives and the product of two $\psi$ fields with first-order derivatives. Here we drop the irrelevant terms, \textit{i.e.}, product of four fields, whose corresponding scaling dimension $\Delta^{(4)} = 2 + \mathrm{number~of~derivatives}$.

\section{TBA solution of the chiral Heisenberg chain}\label{TBA}

In this section, we present the detailed solution of the chiral Heisenberg chain with thermodynamic Bethe ansatz, including the Bethe equation, the thermodynamic limit, and the equation for the free energy.

\subsection{Bethe equations}
The set of BE for the Hamiltonian \cref{hamiltonian} can be written in the form
\begin{equation}
\label{rap}
\left(\frac{x_j+i}{x_j-i}\right)^N=\prod_{l \neq j}\left(\frac{x_j-x_l+2 i}{x_j-x_l-2 i}\right), \quad j=1, \ldots, M.
\end{equation}
where $x_j$ are the rapidities.
The corresponding energy in terms of the rapidities solving \cref{rap} are
\begin{equation} \label{BetheEquation}
E\left(x_1, \ldots, x_M\right)=-\sum_{j=1}^M\left(J - \alpha \partial_{x_j}\right) \frac{2}{x_j^2+1}.
\end{equation}
The solution of \cref{rap} can be  assumed to live on a collection of strings\cite{takahashi1997thermodynamical} and have the following form
\begin{equation}
x_{j} \to x_\mu^{n, j}=x_\mu^n+i(n+1-2 j), \quad j=1, \ldots, n.
\end{equation}
Here $\mu$ labels different strings, $n$ labels the length of the strings, and $j$ labels the position of the string.

\Cref{BetheEquation} then becomes
\ifTwocolumn \begin{widetext} \fi
\begin{equation}
e^N\left(x_\mu^{n, j}\right)=\prod_{(m, \nu) \neq(n, \mu)} e\left(\frac{x_\mu^{n, j}-x_\nu^m}{m-1}\right) e\left(\frac{x_\mu^{n, j}-x_\nu^m}{m+1}\right) \prod_{j^{\prime} \neq j} e\left(\frac{x_\mu^{n, j}-x_\mu^{n, j^{\prime}}}{2}\right),
\end{equation}
where we defined the function $e(x)$ as follows:
\begin{equation}
    e(x) = \frac{x+i}{x-i}.
\end{equation}

This can be further simplified to
\begin{equation}
e^N(x_\mu^n/n)=\prod_{j=1}^ne^N(x_\mu^{n,j})=\prod_{(m,\nu)\neq(n,\mu)}E_{nm}(x_\mu^n-x_\nu^m),
\end{equation}
if we define
\begin{equation}
    E_{nm}(x)\equiv\left\{
    \begin{array}{cc}e(\frac{x}{|n-m|})e^2(\frac{x}{|n-m|+2})e^2(\frac{x}{|n-m|+4})...e^2(\frac{x}{n+m-2})e(\frac{x}{n+m}) \text{ for } n\neq m,\\
    e^2(\frac{x}{2})e^2(\frac{x}{4})...e^2(\frac{x}{2n-2})e(\frac{x}{2n}) \text{ for } n=m.
    \end{array}\right.
\end{equation}

The logarithm form of these equations gives
\begin{equation} \label{logBE}
    \begin{aligned}
        N\theta(x_\mu^n/n)&=2\pi I_\mu^n+\sum_{(m,\nu)\neq(n,\mu)}\Theta_{nm}(x_\mu^n-x_\nu^m),
    \end{aligned}
\end{equation}
where
\begin{equation}
    \Theta_{nm}(x)\equiv\left\{
    \begin{matrix}\theta(\frac{x}{|n-m|})+2\theta(\frac{x}{|n-m|+2})+...+2\theta(\frac{x}{n+m-2})+\theta(\frac{x}{n+m}) \text{ for } n\neq m,\\
    2\theta(\frac{x}{2})+2\theta(\frac{x}{4})+...+2\theta(\frac{x}{2n-2})+\theta(\frac{x}{2n}) \text{ for } n=m,
    \end{matrix}\right.
\end{equation}
and
\begin{equation}
    \theta(x)\equiv2\arctan(x).
\end{equation}

\subsection{Thermodynamic Bethe equation}

In the thermodynamic limit, \cref{logBE} becomes
\begin{equation}
\begin{aligned}
    2\pi\int^{x}\rho_{n}(t)+\rho_{n}^{h}(t)dt=\theta_{n}(x)-\sum_{m=1}^{\infty}\int_{-\infty}^{\infty}\Theta_{nm}(x-y)\rho_{m}(y)dy,
\end{aligned}
\end{equation}
where $\rho_n(x)$ ($\rho_n^h(x)$) is the density of the strings (holes).
Differentiating this equation with respect to $x$ one obtains
\begin{equation} \label{ThermoBetheEquation}
    a_n(x)=\rho_n(x)+\tilde{\rho_n}(x)+\sum_k T_{j k} * \rho_k(x),
\end{equation}
where
\begin{equation} \label{Tnm}
    T_{n m}(x) \equiv\left\{\begin{aligned}
    &a_{|n-m|}(x)+2 a_{|n-m|+2}(x)+2 a_{|n-m|+4}(x)+
    \cdots+2 a_{n+m-2}(x)+a_{n+m}(x), \text { for } n \neq m, \\
    &2 a_2(x)+2 a_4(x)+\cdots+2 a_{2 n-2}(x)+a_{2 n}(x), \text { for } n=m .
\end{aligned}\right.
\end{equation}
\ifTwocolumn \end{widetext} \fi
and we introduced the following notations
\begin{equation}
\begin{aligned}
    a_n(x)&\equiv\frac{1}{\pi}\frac{n}{x^2+n^2}, \\
    a_0(x)&\equiv\delta(x).
\end{aligned}
\end{equation}

The Free energy per site, $f = e - Ts$, can be written as
\begin{equation}\begin{aligned}
e=\frac{J}{4}+\sum_{n=1}^{\infty}\int_{-\infty}^{\infty}g_{n}(x)\rho_{n}(x)dx,
\end{aligned}\end{equation}
where
\begin{equation}\begin{aligned}
g_{n}(x)\equiv -2\pi(J - \alpha \partial_x)a_{n},
\end{aligned}\end{equation}
and
\begin{dmath}
    s=\sum_{n=1}^\infty\int_{-\infty}^\infty\rho_n(x)\ln(1+\frac{\rho_n^h(x)}{\rho_n(x)})
+\rho_n^h(x)\ln(1+\frac{\rho_n(x)}{\rho_n^h(x)})dx.
\end{dmath}
Minimizing the free energy, one obtains the following equation
\begin{dmath}
0 = \delta e-T\delta s \\
=\sum_{n=1}^{\infty}\int dx \left\{\left[g_{n}(x)-T\ln(1+\frac{\rho_{n}^{h}(x)}{\rho_{n}(x)})\right]\delta\rho_{n}(x)
-T\ln(1+\frac{\rho_{n}(x)}{rho_{n}^{h}(x)})\delta\rho_{n}^{h}(x)\right\}.
\end{dmath}
Using \cref{ThermoBetheEquation}, the above equation can be cast in the form
\begin{equation}
\delta\rho_n^h(x)=-\delta\rho_n(x)-\sum_m\int T_{nm}(x-y)\delta\rho_m(y)dy.
\end{equation}
From here, one obtains
\begin{equation} \label{eta_sum}
\ln \eta_n=\frac{g_n}{T}+\sum_{k=1}^{\infty} T_{n k} * \ln \left(1+\eta_k^{-1}\right),
\end{equation}
where we define $\eta_n(x) = \rho_n^h(x) / \rho_n(x)$.

For $n = 1$, it reads
\begin{dmath}
\label{eta1_raw}
\ln\eta_1=\frac{g_1(x)}{T}+a_2*\ln(1+\eta_1^{-1})
+\sum_{j=2}^{\infty}(a_{j-1}+a_{j+1})*\ln(1+\eta_j^{-1}).
\end{dmath}
This can be equivalently rewritten as
\begin{equation} \label{eta1}
\ln(1+\eta_1)=\frac{g_1(x)}{T}+\sum_{l=1}^{\infty}(a_{l-1}+a_{l+1})*\ln(1+\eta_l^{-1}).
\end{equation}
Finally, from \cref{Tnm}, one obtains a recursive relation for $a_0$, $a_1$ and $a_2$ of the form:
\begin{dmath} \label{aT}
a_1*(T_{n-1,m}+T_{n+1,m})-(a_0+a_2)*T_{n,m}
=(\delta_{n-1,m}+\delta_{n+1,m})a_1.
\end{dmath}

Now, combining \cref{eta_sum,aT}, the whole set of recursive relations of $\eta_n$ reads
\ifTwocolumn
    \begin{equation} \label{aeta}
    \begin{aligned}
        (a_{0}+a_{2})*\ln\eta_{1}(x)&=\frac{g_1(x)}{T}+a_{1}*\ln(1+\eta_{2}(x)), \\
        (a_0+a_2)*\ln\eta_n(x)&=a_1*\ln(1+\eta_{n-1}(x))(1+\eta_{n+1}(x)), \\
        n=2,3,...
    \end{aligned}
    \end{equation}
\else
    \begin{equation} \label{aeta}
    \begin{aligned}
        (a_{0}+a_{2})*\ln\eta_{1}(x)&=\frac{g_1(x)}{T}+a_{1}*\ln(1+\eta_{2}(x)), \\
        (a_0+a_2)*\ln\eta_n(x)&=a_1*\ln(1+\eta_{n-1}(x))(1+\eta_{n+1}(x)), \quad n=2,3,...
    \end{aligned}
    \end{equation}
\fi
Substitution of \cref{aeta} into \cref{eta1_raw} yields
\begin{dmath}
0 = a_{n}*\ln\eta_{n+1}
-{a_{n+1}*\ln(1+\eta_{n})-a_{n+2}*\ln(1+\eta_{n+1}^{-1})}
-\sum_{l=n+2}^\infty(a_{l-1}+a_{l+1})*\ln(1+\eta_l^{-1}).
\end{dmath}
Equivalently, one can rewrite $\ln\eta_{n+1}$ in terms of other $\ln\eta_{l}$ as follows:
\begin{dmath}
\ln\eta_{n+1}=a_{1}*\ln\eta_{n}
+a_{2}*\ln(1+\eta_{n+1}^{-1})
+\sum_{l=n+2}^{\infty}(a_{l-n-1}+a_{l-n+1})*\ln(1+\eta_{l}^{-1}).
\end{dmath}
For large $n$, $\ln(1+\eta_n^{-1})\simeq o(n^{-2})$ and therefore:
\begin{equation}
\lim\limits_{n\to\infty}\ln\eta_{n+1}-a_1*\ln\eta_n=0,
\end{equation}
or equivalently
\begin{equation}
\lim_{n\to\infty}\frac{\ln\eta_n}{n}=0.
\end{equation}
So, one can summarize the set of thermodynamic BE as
\begin{equation} \label{eta}
\begin{aligned}
&\ln\eta_{1}(x)=\frac{-2\pi(J - \alpha \partial_x)}{T}s(x)+s*\ln(1+\eta_{2}(x)), \\
&\begin{aligned}\ln\eta_{n}(x)=s*\ln(1+\eta_{n-1}(x))(1+\eta_{n+1}(x)),
\end{aligned} \\
&\operatorname*{lim}_{n\to\infty}\frac{\ln\eta_{n}}{n}=0,
\end{aligned}\end{equation}
where
\begin{equation}
s(x)=\frac{1}{4}\sech(\frac{\pi x}{2}).
\end{equation}

Having written the set of BE in the thermodynamic limit, one can consider the analytical expression for the free energy, which becomes
\begin{equation}
\begin{aligned}
   f&=e-Ts \\
   &=\frac{J}{4}+\sum_{n=1}^{\infty}\int g_{n}\rho_{n}-T[\rho_{n}\ln(1+\eta_{n})+\rho_{n}^{h}\ln(1+\eta_{n}^{-1})]dx. 
\end{aligned}
\end{equation}
Here one can eliminate $\rho_n^h$ using \cref{ThermoBetheEquation}. Thus, the free energy acquires the form:
\ifTwocolumn
    \begin{equation}\begin{aligned}
    f&=\frac{J}{4}-T\sum_{n=1}^{\infty}\int\ln(1+\eta_{n}^{-1})a_{n}(x) \\
    &+\rho_{n}[\ln\eta_{n}-\frac{g_{n}}{T}-T_{nm}*\ln(1+\eta_{m}^{-1})]dx \\
    &=\frac{J}{4}-T\sum_{n=1}^{\infty}\int a_{n}(x)\ln(1+\eta_{n}^{-1}(x))dx.
    \end{aligned}\end{equation}
\else
    \begin{equation}\begin{aligned}
    f&=\frac{J}{4}-T\sum_{n=1}^{\infty}\int\ln(1+\eta_{n}^{-1})a_{n}(x)
    +\rho_{n}[\ln\eta_{n}-\frac{g_{n}}{T}-T_{nm}*\ln(1+\eta_{m}^{-1})]dx \\
    &=\frac{J}{4}-T\sum_{n=1}^{\infty}\int a_{n}(x)\ln(1+\eta_{n}^{-1}(x))dx.
    \end{aligned}\end{equation}
\fi
Now, multiplying \cref{eta1} by $s(x)$ and integrating it over $x$ gives
\begin{dmath}
\int dxs(x)\ln(1+\eta_1) = \frac{1}{T}\int s(x) g_1(x)dx
+ \sum_{l=1}^{\infty}\int a_l(x)\ln(1+\eta_l^{-1}(x))dx.
\end{dmath}
Thus one obtains for the free energy of the model an integral form: 
\begin{equation}
f=-J\left(\ln 2 - \frac{1}{4}\right) - T\int s(x)\ln(1+\eta_{1}(x))dx.
\end{equation}

\subsection{Low temperature limit}

To take the $T \to 0$ limit in the expression for the free energy, a new set of variables are introduced
\begin{equation}
\eta_{n}(x)=\exp\{\epsilon_{n}(x)/T\},n=1\ldots N.
\end{equation}
Then, \cref{eta} acquire the following form:
\ifTwocolumn
    \begin{equation}
    \begin{aligned}
    \epsilon_1(x) &= -2 \pi (J - \alpha \partial_x) s(x)+s * T \ln \left(1+\exp \left(\frac{\epsilon_2(x)}{T}\right)\right) \\
    \epsilon_n(x) &= s * T \ln \left[ \left(1+\exp \left(\frac{\epsilon_{n-1}(x)}{T}\right)\right) \right. \\
    &\left.\left(1+\exp \left(\frac{\epsilon_{n+1}(x)}{T}\right)\right) \right]\\
    & \lim _{n \rightarrow \infty} \frac{\epsilon_n(x)}{n}=0
    \end{aligned}
    \end{equation}
\else
    \begin{equation}
    \begin{aligned}
    \epsilon_1(x) &= -2 \pi (J - \alpha \partial_x) s(x)+s * T \ln \left(1+\exp \left(\frac{\epsilon_2(x)}{T}\right)\right) \\
    \epsilon_n(x) &= s * T \ln \left[ \left(1+\exp \left(\frac{\epsilon_{n-1}(x)}{T}\right)\right)
    \left(1+\exp \left(\frac{\epsilon_{n+1}(x)}{T}\right)\right) \right]\\
    & \lim _{n \rightarrow \infty} \frac{\epsilon_n(x)}{n}=0
    \end{aligned}
    \end{equation}
\fi
Then, the free energy becomes
\begin{equation}
f=-J\left(\ln2-\frac{1}{4}\right)-T\int s(x)\ln(1+\exp(\epsilon_1(x)/T))dx,
\end{equation}
which yields  \cref{freeenergy} of the main text.

\section{Chirality when \texorpdfstring{$\alpha > \alpha_c$}{alpha > alphac}} \label{chiralityLargeAlpha}

The chirality can be computed by taking derivative of the free energy \cref{falpha} with respect to $\alpha$
\ifTwocolumn
    \begin{equation}\begin{aligned}
    \frac{1}{N}\chi =& \frac{df}{d\alpha} \\
    =& -\frac{1}{2}\sum_{nk} (-1)^{k+n} \pi \left(k+\frac{1}{2}\right) \frac{e^{-\pi a\left(k+n+1\right)}G(n)G(k)}{k+n+1} \\
    &+\frac{1}{2}\sum_{nk} (-1)^{k+n} \left[1-\frac{\pi\alpha}{J}\left(k+\
    \frac{1}{2}\right)\right] \\
    &\times e^{-\pi a\left(k+n+1\right)} G(n)G(k) \left(-\pi\right) J\frac{da}{d\alpha}.
    \end{aligned}
    \end{equation}
\else
    \begin{equation}\begin{aligned}
    \frac{1}{N}\chi =& \frac{df}{d\alpha} \\
    =& -\frac{1}{2}\sum_{nk} (-1)^{k+n} \pi \left(k+\frac{1}{2}\right) \frac{e^{-\pi a\left(k+n+1\right)}G(n)G(k)}{k+n+1} \\
    &+\frac{1}{2}\sum_{nk} (-1)^{k+n} \left[1-\frac{\pi\alpha}{J}\left(k+\
    \frac{1}{2}\right)\right]
    e^{-\pi a\left(k+n+1\right)} G(n)G(k) \left(-\pi\right) J\frac{da}{d\alpha}.
    \end{aligned}
    \end{equation}
\fi

The second part disappears after factorizing
\ifTwocolumn
    \begin{equation}
    \begin{aligned}
        \chi_2 \equiv& \frac{1}{2}\sum_{nk} (-1)^{k+m} \left[1-\frac{\pi\alpha}{J}\left(k+\
        \frac{1}{2}\right)\right] \\
        &\times e^{-\pi a\left(k+n+1\right)} G(n)G(k) \left(-\pi\right) J \frac{da}{d\alpha} \\
        =&-\frac{\pi}{2} J \frac{da}{d\alpha}\sum_{n}(-1)^{n}f_{n}\sum_{k}\left[1-\frac{\pi\alpha}{J}(k+\frac{1}{2})\right](-1)^{k}f_{k} \\
        =&0,
    \end{aligned}
    \end{equation}
\else
    \begin{equation}
    \begin{aligned}
        \chi_2 \equiv& \frac{1}{2}\sum_{nk} (-1)^{k+m} \left[1-\frac{\pi\alpha}{J}\left(k+\
        \frac{1}{2}\right)\right] 
        e^{-\pi a\left(k+n+1\right)} G(n)G(k) \left(-\pi\right) J \frac{da}{d\alpha} \\
        =&-\frac{\pi}{2} J \frac{da}{d\alpha}\sum_{n}(-1)^{n}f_{n}\sum_{k}\left[1-\frac{\pi\alpha}{J}(k+\frac{1}{2})\right](-1)^{k}f_{k} \\
        =&0,
    \end{aligned}
    \end{equation}
\fi
where
\begin{equation}
    f_k = e^{-\pi a (k+1/2)}G(k),
\end{equation}
and the last equality holds as a result of \cref{aofalpha}
\begin{equation}
    \sum_{k=0}^\infty(-1)^k\left[1-\frac{\pi\alpha}{J}(k+1/2)\right] f_k=0.
\end{equation}

So the remaining part of the chirality
\begin{equation}\begin{aligned}
    \chi/N&= -\frac12\sum_{nk}(-1)^{k+n}\pi\left(k+\frac12\right)\frac{e^{-\pi a(k+n+1)} G(n)G(k)}{k+n+1}\\
    &=  -\frac14\sum_{nk}(-1)^{k+n}\pi e^{-\pi a(k+n+1)} G(n)G(k) \\
    &=  -\frac\pi4 \left(\sum_{n}(-1)^{n} e^{-\pi a(n+1/2)} G(n)\right)^2,
\end{aligned}\end{equation}
which is the \cref{chiSeries} presented in the main text.

\section{
Quadratic dependence of the ground state energy near the transition point as a function of \texorpdfstring{$\alpha$}{alpha} 
} \label{closedform}

\begin{figure}[t!]
    \centering
    \begin{minipage}{0.55\textwidth}
    \includegraphics[width=\textwidth]{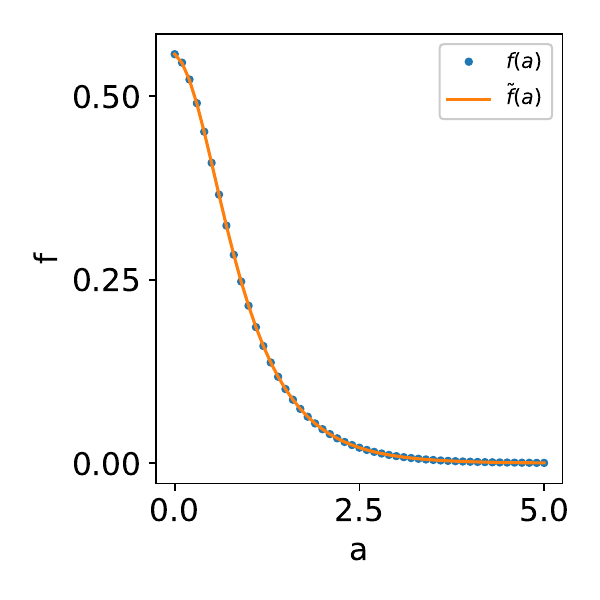}
    \end{minipage}
    \caption{\label{fa}
Precision fitting of the series representation of $f(a)$ (\cref{series_fa}) with a closed form \cref{fitting_fa}.
    }
\end{figure}

We find a closed analytical expression for the free energy near the critical point in this section.
We first define
\begin{equation} \label{series_fa}
    f(a) = \sum_k (-1)^k e^{-\pi a (k+1/2)} G(k),
\end{equation}
which can be fitted to a great precision with a simple function
\begin{equation} \label{fitting_fa}
    \tilde{f}(a) = \left(\frac{2\gamma_0 - \sqrt{\frac{\pi}{e}}}{e^{\pi a}} + \sqrt{\frac{\pi}{e}}\right)\frac{e^{\pi a/2}}{e^{\pi a} + 1},
\end{equation}
as shown in \cref{fa}, where $\gamma_0 = \sum_k (-1)^k G(k) \approx 0.557$.

Then \cref{alphaa} can be written as
\begin{equation}
\begin{aligned}
    \alpha / J &= -\frac{f}{f'} \simeq -\frac{\tilde{f}}{\tilde{f}'} \\
    &= \frac{2}{\pi} + \frac{8 e^{\pi  a} \left(\sqrt{\pi }-\sqrt{e} \gamma_0\right)}{2 \pi  \sqrt{e} \left(3 e^{\pi  a}+1\right) \gamma_0+\pi ^{3/2} \left(-4 e^{\pi  a}+e^{2 \pi  a}-1\right)}.
\end{aligned}
\end{equation}

We are interested in the vicinity of $\alpha = 2/\pi$ and $a~\to~\infty$.
Under such limit,
\begin{equation}
    e^{\pi a} \simeq \frac{8 \left(\sqrt{\pi }-\sqrt{e} \gamma_0\right)}{\pi ^{3/2} (\alpha/J - 2/\pi)}.
\end{equation}

The corresponding chirality
\begin{equation}
\begin{aligned}
    \chi &\simeq -\frac{\pi ^2}{4}  e^{-\pi  a-1} \\
    &\simeq -\frac{\pi ^{7/2}}{32 e \left(\sqrt{\pi }-\sqrt{e} \gamma_0\right)} (\alpha/J - 2/\pi),
\end{aligned}
\end{equation}
and hence the chiral part of the energy per particle becomes
\begin{equation}
\begin{aligned}
    F_{\chi}/J &\simeq \frac{\pi ^2}{4}  e^{-\pi  a-1} \\
    &\simeq -\frac{\pi ^{7/2}}{64 e \left(\sqrt{\pi }-\sqrt{e} \gamma_0\right)} (\alpha / J - 2/\pi)^2 \\
    &\approx -0.370154 (\alpha / J - 2/\pi)^2.
\end{aligned}
\end{equation}
This derivation reproduces \cref{chasym} of the main text. 
\vspace{0.25mm}

\providecommand{\href}[2]{#2}\begingroup\raggedright\endgroup

\bibliographystyle{JHEP}

\end{document}